\documentclass[twocolumn,showpacs,amsmath,amssymb,
superscriptaddress,pre]{revtex4}

\usepackage{amsmath,amssymb,latexsym,graphicx,epsfig}
\usepackage{pstricks,psfrag}
\newpsobject{showgrid}{psgrid}{subgriddiv=1}

\numberwithin{equation}{section}
\begin{document}

\title{Renormalization group
theory for finite-size scaling in extreme statistics
}

\author{G. Gy\"{o}rgyi}
\altaffiliation[Present address:] {Department of Materials Physics,  E\"{o}tv\"{o}s University, Budapest, Hungary}
\email{gyorgyi@glu.elte.hu}
\affiliation{Department of Theoretical Physics, University of Geneva, Geneva,
Switzerland}

\author{N. R. Moloney}
\altaffiliation[Present address:]{Max Planck Institute for
the Physics of Complex Systems, N\"othnitzer Str. 38, 
D-01187}
\email{moloney@pks.mpg.de}
\affiliation{Institute of Theoretical Physics - HAS,
  E\"{o}tv\"{o}s University, Budapest, Hungary}

\author{K. Ozog\'{a}ny}
\email{ozogany@general.elte.hu}
\affiliation{Institute of Theoretical Physics - HAS Research Group,
  E\"{o}tv\"{o}s University, Budapest, Hungary}

\author{Z. R\'{a}cz}
\email{racz@general.elte.hu}
\affiliation{Institute of Theoretical Physics - HAS,
  E\"{o}tv\"{o}s University, Budapest, Hungary}

\author{M. Droz}
\email{Michel.Droz@unige.ch}
\affiliation{Department of Theoretical Physics, University of Geneva, Geneva,
Switzerland}

\date{\today}

\begin{abstract}

We present a renormalization group (RG) approach to explain
universal features of extreme statistics, applied here
to independent, identically distributed
variables. The outlines of the theory have been described in
a previous Letter, the main result being that finite-size
shape corrections to the limit distribution can be obtained
from a linearization of the RG transformation near a
fixed point, leading to the computation of stable perturbations as
eigenfunctions.  Here we show details of the RG theory which exhibit
remarkable similarities to the RG known in statistical physics.
Besides the fixed points explaining universality, and the least
stable eigendirections accounting for convergence rates and
shape corrections, the similarities include
marginally stable perturbations
which turn out to be generic for the Fisher-Tippett-Gumbel class.
Distribution functions containing unstable perturbations are
also considered. We find that, after a transitory divergence, 
they return to the universal fixed line at the same or at a different 
point depending on the type of perturbation. 
\end{abstract}
\pacs{{05.40.-a, 02.50.-r, 05.45.Tp}}

\maketitle

%\tableofcontents

%%%%%%%%%%%%%%%%%%%%%%%%%%%%%%%%%%%%%%%%%%%%%%%%%%%%%%%%%%%%%%%%%%%%%%%%%%%%
\section{Introduction}
\label{S:intro}
%%%%%%%%%%%%%%%%%%%%%%%%%%%%%%%%%%%%%%%%%%%%%%%%%%%%%%%%%%%%%%%%%%%%%%%%%%

Using present data-acquisition methods, it has become
possible to build enormous databases in practically all
fields of science. The large size of these databases
make their analysis rather intricate, but questions can now
be posed which were previously unanswerable. In particular,
quantitative studies of extreme events have become feasible.
Accordingly, extreme value statistics (EVS) has
found its way from mathematics \cite{FisherTippett:1928,Gumbel:1958}
to various disciplines, foremost among them those fields where
the extremes involve great risks as in
engineering\cite{Weibull:1951}, finance \cite{EmbrechtETAL:1997},
hydrology \cite{KatzParlangeNaveau:2002}, meteorology \cite{StorchZwiers:2002},
geology \cite{GutenbergRichter:1944}. In physics, extreme statistics was
relatively dormant until the last decade when
a number of studies appeared in relation to spin glasses \cite{BouchaudMezard:1997},
interface-- and landscape--fluctuations \cite{RaychaudhuriETAL:2001, AntalETAL:2001,
GyorgyiETAL:2003, LeDoussalMonthus:2003, MajumdarComtet:2004, MajumdarComtet:2005,
BolechRosso:2004,GucluKorniss:2004}, random fragmentation
\cite{KrapivskyMajumdar:2000}, level-density in ideal quantum gases
\cite{Comtet:2007}, atmospheric time series \cite{EichnerETAL:2006}, etc.

The above studies revealed that an important problem with the
quantitative application of EVS is the
slow convergence in the
size $N$ of the data set whose extremum is sought. A well known
example is that the
Fisher-Tippett-Gumbel limit distribution
for independent, identically distributed (i.i.d.)
variables is typically approached logarithmically
slowly, i.e. proportionally to $1/\ln N$. This convergence problem
is not cured even by the present size of the
databases, thus studies of finite-size corrections
are needed for a thorough analysis of EVS. While the
mathematical literature contains results on the rate
of convergence to the universal limit
distributions \cite{DeHaanResnick:1996,DeHaanStadm:1996}, and
particular cases for correlated variables
have also been studied in physics \cite{SchehrMajumdar:2006},
to our knowledge, finite-size effects have apparently been ignored
in most empirical data analysis.

In order to develop a readily usable theory of finite-size
corrections giving not only the amplitude of the corrections
but also providing shape corrections to the limit
distribution, we recently introduced a
renormalization group (RG) approach for the case of
i.i.d. variables \cite{GyorgyiETAL:2008}. 
The concept of using RG in probability theory 
has been pioneered by Jona-Lasinio \cite{JonaLasinio:2001}
in connection with critical phenomena. In the context of EVS, 
the real-space RG methods have been applied to random 
landscapes \cite{LeDoussalMonthus:2003,SchehrLeDoussal:2009}.
In our case, it turned out that
the RG equation
is just a physicist's reformulation and a simple generalization
of the original fixed point equation of Fisher and Tippett
\cite{FisherTippett:1928}. Within this approach, the fixed
point condition yields the 
known one-parameter ($\gamma$) family of solutions
which provides the traditional classification of EVS limit
distributions (Fisher-Tippett-Gumbel, -Fr\'echet,
-Weibull). The RG approach is instrumental in determining the
finite-size corrections as well, since they emerge from the
action of the RG on the neighborhood of the fixed point.
We found that the resulting eigenvalue problem and the emerging
eigenfunctions represent the generic
leading finite-size correction functions. An important
outcome of the general considerations is that the
first order shape corrections display universality.
Namely, within a class of limit distributions characterized
by $\gamma$, the correction is determined by a single 
additional parameter $\gamma^\prime$. The $\gamma$ is long known 
to be related to the leading asymptote of 
the parent distribution, while $\gamma'$, emerging as an eigenvalue
index from RG, is determined by the next-to-leading asymptote
of the parent distribution which is in accordance 
with the mathematical results 
\cite{DeHaanResnick:1996,DeHaanStadm:1996}.
Note that our RG framework also provides a simple
example in addition to a series of known exact RG schemes 
\cite{JonaLasinio:2001,Ma:1973,Hilhorst:1978,McKay:1982,Schuster:2005}.

The above RG picture and the corresponding results have been
briefly presented in a recent Letter \cite{GyorgyiETAL:2008}.
Here, we offer an extended and more detailed view on our work
with new aspects relating to the nonlinear neighborhood of the 
fixed line including nonlinear manifolds.
In particular, a detailed derivation of the RG picture is presented
in Sec.~\ref{S:iid-fp} where
the fixed line of the RG transformation, parametrized by $\gamma$,
is linked to the EVS limit distributions.
Next, linear perturbations around a point on the
fixed line are considered.
The eigenvalue problem is solved (Sec.~\ref{S:iid-lin}), giving
rise to a new parameter $\gamma^\prime$.
The general aspects of the theory are developed further in
Sec.~\ref{S:iid-nlrg} where the
RG transformation beyond the linear neighborhood
of the fixed point is formulated to treat the case of 
marginal eigenfunctions within the Fisher-Tippett-Gumbel class. 
The practical consequences are discussed by connecting to
the finite size properties in typical EVS problems and showing
that the parameter $\gamma^\prime$ is the finite-size scaling 
exponent (Sec.~\ref{S:iid-lin-sd}). Furthermore, the relationship
between $\gamma^\prime$ and the parent distributions is determined, 
and we analyze in detail a family of
parents with generalized exponential asymptote frequently occuring
in applications.
The question of unstable perturbations and the invariant
manifolds associated with them are 
examined in Sec.~\ref{S:iid-unst}, and they are illustrated on
a series of examples.
Finally, the conclusion and  possible avenues for
further application of the RG scheme to EVS problems are
found in Sec.~\ref{S:end}.

%%%%%%%%%%%%%%%%%%%%%%%%%%%%%%%%%%%%%%%%%%%%%%%%%%%%%%%%%%%%%%%%%%%%%%%%%%%%
\section{Renormalization group transformation and its fixed line}
\label{S:iid-fp}
%%%%%%%%%%%%%%%%%%%%%%%%%%%%%%%%%%%%%%%%%%%%%%%%%%%%%%%%%%%%%%%%%%%%%%%%%%%%

In this section we recuperate known basic properties of EVS and
explain the emergence of limit distributions within the framework
of RG  theory.  This is essentially a reformulation of the
original proposition of the seminal work by Fisher and Tippett
\cite{FisherTippett:1928}, which we will further develop in subsequent sections.

The basic question of EVS is the limit behavior of the
distribution of extremal values of random variables.
In what follows we shall restrict our attention to
maxima, i.e., the distribution of the largest variable,
but by a sign change
this is equivalent to the problem of minima.
Suppose that we have a
``batch'' of random variables $z_1,z_2,\dots , z_N$,
and we ask about the statistical properties of the largest
$z=\max{\left\lbrace z_1,z_2,\dots , z_N\right\rbrace }$. The question has a
simple answer, if the original variables $z_i$ were independent, identically
distributed, each according to the so called parent probability density
$\rho(z)$.  The simplicity is apparent when using
the integrated (cumulative) distribution
\begin{equation}
  \mu(z) = \int_{-\infty}^z \!\rho(\bar z)d\bar z,
\end{equation}
i.e. the probability that the  variable is not larger than $z$ (we
assume here and later nonsingular densities, and differentiability if necessary).  Then the probability that the maximum is not
exceeding $z$ is given by the joint
probability that all variables do not exceed $z$, hence the integrated
distribution of the maximum of batch of $N$ variables, the extreme value
distribution (EVD) is
\begin{equation}
 M_N^{\max}(z) = \mu^N(z).\label{eq:mN-def}
\end{equation}

Since the integrated parent distribution $\mu(z)$ is less than
one whenever $z$ is within the support of the parent, the above
quantity will converge to zero
there for any fixed $z$.   However, we can ask whether there is a
linear change of variable so that for large $N$ the EVD is restored
to a nondegenerate function in the large $N$ limit.
Specifically, we ask about the existence of parameters $a_N,b_N$
such that
\begin{equation}
 M_N(x)=M_N^{\max}(a_Nx+b_N) \to M(x),\label{eq:mN-limit}
\end{equation}
where the resulting integrated distribution represents the extreme value limit
distribution. It is related to the limit density as
$P(x)=M^\prime (x)$.

Note that the parameters $a_N,b_N$ are not unique, if one choice produces a
nondegenerate limit EVD then adding constants to either will also yield another one. This ambiguity can be eliminated by a
standardization convention, for example
\begin{equation}
M(0)=P(0)=1/e. \label{eq:mp-norm}
\end{equation}
We can require this condition also at finite stages in the limit
\eqref{eq:mN-limit} for the maximum distribution $M_N(x)$, which specifies
uniquely the parameters $a_N,b_N$.

Now let us turn to the invariance condition, which follows from the 
fact that
$M(x)$ is a limit distribution of EVS. The reason behind this invariance 
is that if $p$ variables come from a parent that 
is a limit EVD, then their maximum will be
also distributed according to the same limit EVD.  
Indeed, conceiving each of the variables as themselves 
representing  maxima of some other large collection
of variables, the distribution of their overall maximum will obviously be again
the same limit EVD.  Since these statements are valid up
to a linear change of variable, for the limit EVD we obtain
\begin{eqnarray}
 M(x) = M^p(a(p)x+b(p)),
\label{eq:fp-m}
\end{eqnarray}
with appropriate scale- and shift paramenters, $a(p)$ and $b(p)$.
While here $p$ should be a positive integer, we can easily
convince ourselves that it can be any positive rational.
 Indeed, considering another $N'=Np$, where  $p$ is fixed
during the large size limit, and making use of
\begin{equation}
 M_N(x) \approx M_{Np}(x),
\end{equation}
we obtain by (\ref{eq:mN-def},\ref{eq:mN-limit}) again \eqref{eq:fp-m}.
Then $p$ can in fact be continued to positive real numbers.
The parameters $a(p),b(p)$ in
\eqref{eq:fp-m}  should be consistent  with the standardization
\eqref{eq:mp-norm}.
In sum, condition \eqref{eq:fp-m} should be
understood such that if $M(x)$ is a
limit distribution then it should solve \eqref{eq:fp-m} for  any $p$ with
parameters ensuring \eqref{eq:mp-norm}.   This is known in the mathematical
literature as the condition of  max-stability, appearing first 
in the classic  paper by Fisher and Tippett \cite{FisherTippett:1928} 
and now an exercise in textbooks on EVS.

In order to go later beyond the invariance condition,
following our short paper
\cite{GyorgyiETAL:2008}, we will identify the RG transformation of
an integrated distribution $\mu(x)$ by the following operation
\begin{align}
 [\hat{R}_p \mu](x) &=  \mu^p(a(p)x+b(p)).\label{eq:def-rg}
\end{align}
One can easily see that the standardization of the type
\eqref{eq:mp-norm} for $[\hat{R}_p \mu](x)$ requires
\begin{subequations}
\begin{align}
 b(p) &= \mu^{-1}(e^{-1/p}),\label{eq:stand-b}\\
 a(p) &= p \frac{d\, b(p)}{dp},\label{eq:stand-a}
\end{align} \label{eq:stand} \end{subequations}
\!\!\!a condition forming part of the RG operation.  Note that traditional
RG used in critical phenomena is in fact a semigroup since,
due to the elimination of degrees of freedom,
the inverse element is missing.
In our case, however, the transformations
form a group, as the inverse element of a
transformation with parameter $p$ has parameter $1/p$. Obviously,  $p<1$
corresponds to the decrease of the batch size, but the above
RG transformation is still meaningful.

We should note that, although there are differences,
the EVS version of the RG transformation
[eqs.(\ref{eq:def-rg},\ref{eq:stand})] has the spirit of the
traditional RG. Namely, the raising to power $p>1$
and shifting by $b(p)$ decreases the weight of the small argument part
of the parent distribution
(eliminating the irrelevant degrees of freedom), while rescaling
by $a(p)$ is analogous to the rescaling of
the remaining degrees of freedom.

Within RG theory,  the invariance condition \eqref{eq:fp-m} becomes a
fixed point relation
\begin{equation}
 [\hat{R}_p M](x) =  M(x).\label{eq:rgfp}
\end{equation}
Concerning terminology, we mention that while in mathematical texts
max-stability is essentially the fixed point relation, in the RG framework
stability is  characterized by the response to small deviations
from the fixed point, to be investigated later in the paper.

One can easily solve \eqref{eq:rgfp} by introducing the
function $f(x)$ as
\begin{equation}
M(x)=e^{-e^{-f(x)}},
\label{eq:m-f}
\end{equation}
where standardization  \eqref{eq:mp-norm} requires
\begin{equation}
 f(0)=0,\, \ \ f'(0)=1, \label{eq:f-norm}
\end{equation}
furthermore, from \eqref{eq:stand} the parameters can be
 written in terms of $f$ as
\begin{subequations}
\begin{align}
b(p) &= f^{-1}(\ln p), \label{eq:stand-b0} \\
a(p) &= 1/f'(b(p)). \label{eq:stand-a0}
\end{align}
\label{eq:stand0}
\end{subequations}
The equation for $f$ can be obtained from (\ref{eq:rgfp}), that is,
writing \eqref{eq:fp-m} as
\begin{align}
 f(x) &= f(a(p)x+b(p))-\ln p \label{eq:fp-f}
\end{align}
whence
\begin{align}
 f'(x) &= a(p)f'(a(p) x+b(p) ).
\label{eq:fp-fder}\end{align}
It follows that the sought solution for $f'$,
independent of $p$, is the reciprocal of a linear function. The only form consistent with (\ref{eq:f-norm})
allows one free parameter, $\gamma$, yielding
\begin{align}
 f'(x;\gamma) &= \frac{1}{1+\gamma x},
\end{align}
where here and later $'$ means differentiation by $x$, and
$ b(p) = \gamma^{-1}(a(p)-1)$ should hold.
Again using (\ref{eq:f-norm}) we get by integration the one-parameter family
\begin{equation}
  f(x;\gamma) = \gamma^{-1} \ln (1+\gamma x),\label{eq:fp-fab1}
\end{equation}
and \eqref{eq:stand0} requires
\begin{subequations}
\begin{align}
a(p,\gamma) &=p^\gamma,\label{eq:fp-fa2}\\
b(p,\gamma) &=\gamma^{-1}(p^\gamma-1).
\end{align}
\label{eq:fp-fab2}
 \end{subequations}
Equation (\ref{eq:fp-fab1}) gives with (\ref{eq:m-f})  the well known
generalized extreme value distribution and the respective density
\begin{subequations}
\begin{align}
M(x;\gamma)&=e^{-(1+\gamma x)^{-1/\gamma}}\label{eq:gevd}\\
P(x;\gamma)&=(1+\gamma x)^{-1/\gamma-1}e^{-(1+\gamma
x)^{-1/\gamma}},\label{eq:gevpdf}
\end{align} \label{eq:gev}
\end{subequations}
\!\!with support extending to the range $1+\gamma x\geq 0$.  
This one-parameter family constitutes a 
fixed line of the RG transformation.
The cases  $\gamma<0,\, =0,\ >0$ were studied separately already by Fisher 
and Tippett \cite{FisherTippett:1928}, and correspond to the traditional
categorization into Weibull, Gumbel, and Fr\'echet universality classes,
respectively.  The $\gamma<0$ case gives the limit distribution of integrated
parent distributions approaching $1$ like a power $-1/\gamma$ at a finite upper
border, $\gamma=0$ corresponds to faster than power asymptote at either a finite
limit or at infinity, and $\gamma>0$ represents parent distributions approaching
$1$ at infinity with a power $-1/\gamma$.  These three cases will be referred to by the abbreviations FTW, FTG, and FTF, respectively.

%%%%%%%%%%%%%%%%%%%%%%%%%%%%%%%%%%%%%%%%%%%%%%%%%%%%%%%%%%%%%%%%%%%%%%%%%%%%
\section{Eigenvalue problem near the fixed point}
\label{S:iid-lin}
%%%%%%%%%%%%%%%%%%%%%%%%%%%%%%%%%%%%%%%%%%%%%%%%%%%%%%%%%%%%%%%%%%%%%%%%%%%

%%%%%%%%%%%%%%%%%%%%%%%%%%%%%%%%%%%%%%%%%%%%%%%%%%%%%%%%%%%%%%%%%%%%%%%%%%%
\subsection{Perturbation about the fixed point}
\label{S:iid-lin-pert}
%%%%%%%%%%%%%%%%%%%%%%%%%%%%%%%%%%%%%%%%%%%%%%%%%%%%%%%%%%%%%%%%%%%%%%%%%%%

While the above fixed point condition is a textbook example of deriving the
limit distributions of EVS, its perturbation has not been studied.  Nevertheless, such perturbations  are important since they naturally give rise to finite-size corrections.  Before elucidating the connection to finite-size scaling, let us consider the function space about the fixed point and its transformations under the RG operation represented by the right-hand-side of (\ref{eq:fp-m}). Introduce the perturbed distribution function
\begin{equation}
 M(x;\gamma,\epsilon) = M(x+\epsilon \psi(x);\gamma),
\label{eq:meps}\end{equation}
where $M(x;\gamma)$ is assumed to satisfy the fixed point condition.  We
insert the perturbation into the argument since this leads to a simpler
formulation later. Alternatively, one can also define
\begin{equation}
M(x;\gamma,\epsilon) =  M(x;\gamma)+\epsilon P(x;\gamma)\psi(x).
\label{eq:peps}
\end{equation}
The two formulas are equivalent to linear order in $\epsilon$.  On the other
hand, for a given finite $\epsilon$ the monotonicity condition for the
integrated distribution $M(x;\gamma,\epsilon)$ in $x$ may impose different
restrictions on $\psi(x)$.   Generically, both formulas can be understood as
approximations for small $\epsilon$ of some distribution with general $\epsilon$
dependence, sufficient for the present purposes.  If we consider finite
perturbations, that will be explicitly stated and the monotonicity condition
discussed.

The standardization condition of the type \eqref{eq:mp-norm} for both of the
above integrated distribution functions is satisfied by
\begin{equation}
\psi(0)=0,\, \psi'(0)=0,\ \psi''(0)=-1.
\label{eq:psi}\end{equation}
The third condition sets the sign and scale of $\epsilon$, the minus sign is
just a convention which will be justified later.  Note that in our short
communication \cite{GyorgyiETAL:2008} we used $\psi''(0)=1$, so in comparisons a
sign change should be understood.

%%%%%%%%%%%%%%%%%%%%%%%%%%%%%%%%%%%%%%%%%%%%%%%%%%%%%%%%%%%%%%%%%%%%%%%%%%%
\subsection{Eigenvalues and eigenfunctions}
\label{S:iid-lin-ev}
%%%%%%%%%%%%%%%%%%%%%%%%%%%%%%%%%%%%%%%%%%%%%%%%%%%%%%%%%%%%%%%%%%%%%%%%%%%

Now we shall study the eigenvalue problem of the linearized RG transformation
around the fixed point, specifically, when the RG transformation changes the
magnitude of the perturbation while leaving its functional form invariant.  In
other words, we are seeking the linear part of the invariant manifold of the RG
emanating from the fixed point.

To make notation simpler in this section, we often omit the $\gamma$ and $p$
arguments, when they are obviously implied, e.g., we use $M(x;\epsilon)$ for
$M(x;\gamma,\epsilon)$, etc. We start out from \eqref{eq:meps}, apply the RG
transformation, and assume that the result is of the same functional form with
only the perturbation parameter changing
\begin{equation}
[\hat R_p M(\epsilon)](x)= M^p(a(\epsilon)
x+b(\epsilon);\epsilon)=M(x;\epsilon'),
\label{eq:rg-lin}
\end{equation}
where linearization in $\epsilon$ and $\epsilon'$ is understood.   The
transformation parameters are also expanded
\begin{subequations}
\begin{align}
 a(\epsilon) &= a+\epsilon a_1,\\
 b(\epsilon) &= b + \epsilon b_1,
\end{align}
\end{subequations}
with the zeroth order values $a,b$ taken from (\ref{eq:fp-fab2}) and the first
order corrections $a_1,\, b_1$ to be determined from the standardization
condition
\eqref{eq:stand} with $\mu(x)=M(x;\epsilon)$.  Using \eqref{eq:meps} and
(\ref{eq:m-f}) in these conditions we
straightforwardly get
\begin{subequations}
\begin{align}
 b_1 &= - \psi(b) \\
 a_1 &= - \psi'(b) a.
\end{align}
\label{eq:stand1}
\end{subequations}
\!\!What is left is to determine the possible values of $\epsilon'$ and the
associated invariant function $\psi(x)$.

We now use (\ref{eq:fp-fab1}) and (\ref{eq:m-f}), expand to linear order in
$\epsilon$ and $\epsilon'$ in the second exponential, and introduce their ratio
as eigenvalue
\begin{equation}
\lambda=\epsilon'/\epsilon.
\label{eq:lambda}
\end{equation}
Then we obtain
\begin{equation}
\begin{split}
\lambda a \psi(x) &= \psi(ax+b)+a_1x+b_1 \\
&=\psi(ax+b)-\psi(b)-a\psi'(b)x.
\end{split}
\label{eq:rg1}\end{equation}
We used \eqref{eq:stand1} to obtain the second line, which makes
it obvious that
the r\^ ole of the constants $a_1, b_1$ is to ensure the first
two conditions in \eqref{eq:psi}.

By differentiation we can eliminate the linear part
\begin{equation}
(\lambda/a) \psi''(x) = \psi''(ax+b)
\end{equation}
We remind the reader that $a,b$ are given in (\ref{eq:fp-fab2}) and change
continuously with $p$, furthermore, we recall that
\begin{equation}
 1+\gamma (ax+b) = a(1+\gamma x).
\end{equation}
Thus the $p$ independent solution is
\begin{align}
\psi''(x) &= -(1+\gamma x)^\alpha,
\label{psi}
\end{align}
where the coefficient $-1$ follows from the third
condition in \eqref{eq:psi}, and the eigenvalue is obtained as
\begin{align}
 \lambda &= a^{\alpha+1} = p^{\gamma(\alpha+1)},
\end{align}
where we also used \eqref{eq:fp-fa2}.  The parameter $\alpha$ labels the
eigenfunctions, but it is more convenient to introduce
$\gamma'=\gamma(\alpha+1)$, which gives
\begin{align}
\psi^{\prime\prime}(x;\gamma,\gamma') &=  - (1+\gamma x)^{\gamma'/\gamma-1},
\label{eq:psi-2der}\\
\lambda(\gamma')&=p^{\gamma'}.
\label{eq:evalue}
\end{align}
Linear stability of the fixed point for $p>1$ means that  $\lambda<1$, then
$\gamma'<0$, the marginal case is $\lambda=1, \gamma'=0$, while $\lambda>1,
\gamma'>0$ characterizes unstable directions.

The eigenfunction is obtained by
integrating eq.(\ref{psi}) while satisfying the conditions
(\ref{eq:psi})
\begin{equation}
\begin{split}
\psi(x;\gamma,\gamma') &= - \int_0^x dy \int_0^y dz \, (1+\gamma
z)^{\gamma'/\gamma-1} \\
&= \frac{1+ (\gamma'+\gamma)x   -
(1+\gamma x)^{\gamma'/\gamma+1}}{\gamma'(\gamma'+\gamma)}.
\end{split}
\label{eq:psi-rg}
\end{equation}
Note that for $\gamma=\gamma'$ (\ref{eq:psi-rg}) simplifies to
\begin{equation}
\psi(x;\gamma,\gamma) = - \frac{x^2}{2}
\label{eq:psi-quad}
\end{equation}
irrespective of the value of $\gamma$.  A function equivalent to
(\ref{eq:psi-rg}) has been found in
\cite{DeHaanResnick:1996,GomesDeHaan:1999}
when studying
convergence to the limit distribution, after starting out with parent
distributions in the appropriate domain of attraction.

Knowing $\psi $, a substitution in
(\ref{eq:stand1}) yields the corrections to the parameters as
\begin{subequations}
\begin{align}
a_1&=\frac{p^\gamma(p^{\gamma'}-1)}{\gamma'},\label{eq:rg1-ab} \\
b_1&=\frac{p^{\gamma+\gamma'}}{\gamma'(\gamma+\gamma')} -
\frac{p^\gamma}{\gamma\gamma'} + \frac{1}{\gamma(\gamma+\gamma')}.
\label{eq:rg2-ab}
\end{align}
\end{subequations}

Summing up this section, we found using the RG theory that, for each fixed point specified by a
$\gamma$, the linearized RG transformation
yields a one-parameter family of eigenfunctions
$\psi(x;\gamma,\gamma')$.  The r\^ole of the index  $\gamma'$ will be
elucidated below in Sec.\ref{S:iid-lin-sd}.

%%%%%%%%%%%%%%%%%%%%%%%%%%%%%%%%%%%%%%%%%%%%%%%%%%%%%%%%%%%%%%%%%%%%%%%%%%%
\subsection{Marginal eigenfunctions}
\label{S:iid-lin-mg}
%%%%%%%%%%%%%%%%%%%%%%%%%%%%%%%%%%%%%%%%%%%%%%%%%%%%%%%%%%%%%%%%%%%%%%%%%%%

We can immediately surmise that marginal eigenfunctions should appear
on the line of fixed points $f(x;\gamma)$ \eqref{eq:fp-fab1}. Since two fixed point
functions stay the same upon the action of the RG
transformation, their difference is
also invariant with respect to the RG.  If, in particular, the difference
function was small, i.e., the two fixed points were differing
by an infinitesimal increment
in $\gamma$, then the difference function can be considered
as a linear perturbation about either fixed point and, based on
its invariance upon the RG,
it should be  proportional to the marginal eigenfunction with eigenvalue
$\lambda=1$.   By \eqref{eq:evalue} this implies $\gamma'=0$,
whence we conclude
that $\psi(x;\gamma,0)$ should emerge as an eigenfunction.

The above reasoning is confirmed by a simple calculation.  First we display the
marginal eigenfunction from \eqref{eq:psi-rg}
\begin{equation}
\begin{split}
\psi(x;\gamma,0)
&= \gamma^{-2}\left[ \gamma x   -  (1+\gamma x) \ln (1+\gamma x)\right]
.
\end{split}
\label{eq:psi-marg}
\end{equation}
On the other hand, let us invoke the fixed point distribution
\eqref{eq:gevd} and expand to leading order in
$\gamma-\gamma_0$ about some nearby $\gamma_0$.  It is easy to see that
\begin{equation}
 \frac{\partial}{\partial \gamma}  M(x;\gamma)  = P(x;\gamma)\,
\psi(x;\gamma,0),\label{eq:evd-der-gamma}
\end{equation}
with the definition \eqref{eq:psi-marg}, so finally we obtain for the perturbed
fixed point
\begin{subequations}
\begin{align}
M(x;\gamma) &\approx M(x;\gamma_0) + \epsilon P(x;\gamma_0)\,
\psi(x,\gamma_0,0),
\label{eq:evd-exp}\\
 \epsilon &= \gamma - \gamma_0 . \label{eq:eps-lambda}
 \end{align}
\label{eq:evd-marginal}
\end{subequations}
A comparison with \eqref{eq:peps} shows that indeed the difference of two
infinitesimally close fixed point distributions is given by the marginal
eigenfunction.   This is interesting also from the viewpoint that one can
determine the marginal function without solving the eigenvalue problem, by
differentiation in terms of $\gamma$.   Presumably for non-marginal functions it
remains necessary to consider the traditional eigenvalue equation.

As a side-remark, the scale and sign of the eigenfunctions \eqref{eq:psi-rg}
were defined by the requirement \eqref{eq:psi}.   This had the consequence that
the perturbation formula \eqref{eq:evd-marginal} has as small parameter
$\epsilon$ just the increment in $\gamma$.  Any other convention for the
curvature at the origin in \eqref{eq:psi} would have introduced an extra factor
in the perturbation in \eqref{eq:evd-exp}, so the prescription
\eqref{eq:psi} indeed provides the most natural scale and sign
of the eigenfunctions.

%%%%%%%%%%%%%%%%%%%%%%%%%%%%%%%%%%%%%%%%%%%%%%%%%%%%%%%%%%%%%%%%%%%%%%%%%%%%
\section{Beyond linearized RG in the marginal case}
\label{S:iid-nlrg}
%%%%%%%%%%%%%%%%%%%%%%%%%%%%%%%%%%%%%%%%%%%%%%%%%%%%%%%%%%%%%%%%%%%%%%%%%%%%

Here we shall deal with the situation when $\gamma'=0$, i.e., from
\eqref{eq:evalue} the eigenvalue $\lambda=1$, so within linear theory the size of the perturbation does not change under the action 
of the RG transformation.
That is, the perturbation with such an eigenfunction is marginal, and we have to extend the RG formulation to terms nonlinear in $\epsilon$.

For the fixed point with nonzero $\gamma$, when the parent distribution has a
power asymptote, the case of $\gamma'=0$ seems to be atypical, because
$\psi(x;\gamma,0)$ has a logarithmic singularity as shown in
\eqref{eq:psi-marg}.  On the other hand, in the
FTG class with $\gamma=0$ the case $\gamma'=0$ can be considered as generic,
because  the correction function  $\psi(x;0,0)=-x^2/2$ represents the natural
analytic perturbation in (\ref{eq:meps}).  We shall thus limit 
the RG study in the nonlinear neighborhood of the fixed line to the FTG
case. Consider the most natural extension of (\ref{eq:meps}) with
$\gamma=\gamma'=0$ (these arguments will be omitted in this section) as
\begin{equation}
 M(x;\epsilon) = M(x - \epsilon x^2/2 + \epsilon^2 \psi_2(x)),
\label{eq:meps-nl}\end{equation}
where $M(x)=e^{-e^{-x}}$ and we require
\begin{equation}
 \psi_2(0)=\psi_2^\prime(0)=\psi_2^{\prime\prime}(0)=0\, ,\label{eq:psi2-cond}
\end{equation}
where the first two equations follow from the 
standardization condition \eqref{eq:mp-norm} while the third one 
sets the scale of $\epsilon$.

The RG transformation is again  given by (\ref{eq:rg-lin})
where the right-hand-side should be expanded to second order in $\epsilon$
and
\begin{subequations}  \label{eq:ab-nonlin}
\begin{align}
a(\epsilon )&= 1+\epsilon \ln p +\epsilon^2 a_2,\\
b(\epsilon )&= \ln p + \frac \epsilon 2 \ln^2 p +\epsilon^2 b_2\, .
\end{align} \end{subequations}
Here the terms up to linear order in $\epsilon$ were taken from the  $\gamma,\gamma'\to 0$ limit of (\ref{eq:fp-fab2}).
Since by \eqref{eq:evalue} the eigenvalue is now $\lambda= 1$, i.e., the
eigenfunction is marginal to linear order, we assume an analytic extension of
$\epsilon'=\lambda\epsilon$ as
\begin{equation}
 \epsilon'=\epsilon - r\epsilon^2. \label{eq:marginality}
\end{equation}
Nonlinear stability then requires $|\epsilon'| < |\epsilon|$,  
thus $r\epsilon >0$.  
If $r\epsilon<0$ then we have an unstable direction.

Straightforward calculation based on the RG transformation of \eqref{eq:rg-lin}
together with (\ref{eq:ab-nonlin},\ref{eq:marginality})   yields
\begin{equation}
 \psi_2^{\prime\prime}(x+\ln p) = \psi_2^{\prime\prime}(x)+2\ln p +r.
\end{equation}
The $p$-independent solution $\psi$ of this equation is linear
\begin{align}
\psi_2^{\prime\prime}(x)&=sx,
\end{align}
where $s$ is arbitrary, and $r$ is determined as
\begin{align}
r&=(s-2)\ln p.
\end{align}
Hence
\begin{equation}
\epsilon'=\epsilon -\epsilon^2 (s-2)\ln p,
\label{eq:eps-nl}\end{equation}
and using (\ref{eq:psi2-cond}) we obtain
\begin{equation}
\psi_2(x;s) = \frac s6 x^3.
\label{eq:psi-cub}\end{equation}
Thus we find that in this order a new additional parameter $s$ arises.
The corrections to the transformation  parameters can also be calculated as
\begin{align}
a_2&= \frac{3-s}{2} \ln^2p,\\
b_2 &= \frac{3-s}{6} \ln^3p.
\end{align}

Nonanalytic corrections dominating the quadratic in $\epsilon$ 
can studied by the ansatz
\begin{equation}
 \epsilon'=\epsilon - r|\epsilon|^\xi,\ \ \ 1<\xi<2.\label{eq:eps-xi}
\end{equation}
Carrying out an analysis similar to the above one, we again obtain the cubic
eigenfunction (\ref{eq:psi-cub}) but now $r=s\ln p$.

Next, the above results will be used to obtain 
practicable finite-size expressions in EVS.

%%%%%%%%%%%%%%%%%%%%%%%%%%%%%%%%%%%%%%%%%%%%%%%%%%%%%%%%%%%%%%%%%%%%%%%%%%%
\section{Finite size scaling}
\label{S:iid-lin-sd}
%%%%%%%%%%%%%%%%%%%%%%%%%%%%%%%%%%%%%%%%%%%%%%%%%%%%%%%%%%%%%%%%%%%%%%%%%%%

So far we have been studying the abstract space of distribution
functions under the action of the RG transformation.
Below we explain how the
results on the eigenfunctions near a fixed point can be translated
to the size dependence of the convergence to the limiting EVD
starting out from a particular parent distribution.

%%%%%%%%%%%%%%%%%%%%%%%%%%%%%%%%%%%%%%%%%%%%%%%%%%%%%%%%%%%%%%%%%%%%%%%%%%%
\subsection{The finite-size exponent $\gamma'$}
\label{sS:iid-lin-sd}
%%%%%%%%%%%%%%%%%%%%%%%%%%%%%%%%%%%%%%%%%%%%%%%%%%%%%%%%%%%%%%%%%%%%%%%%%%%

Consider the basic situation described in Sec.\ \ref{S:iid-fp},
namely, the EVD $M_N(x)$ approaches one of the fixed points
$M(x;\gamma)$ for a diverging batch size $N$.
Then it is natural to assume
that the convergence for large $N$ occurs along a dominant
eigenfunction, approximated by the form
\eqref{eq:meps}, with some $\epsilon=\epsilon_N$.
Further $p$ iterates of the
extreme value distribution by the RG transformation
corresponds to increasing the
batch size to $pN$, and so, according to (\ref{eq:rg-lin}), it
generates a new perturbation parameter
\begin{equation}
 \epsilon'=\epsilon_{pN}=\lambda_{\gamma'} \epsilon_N.
\label{eq:eps-pn}\end{equation}
Using now $\lambda (\gamma^\prime)=p^{\gamma^\prime}$
(\ref{eq:evalue}) and assuming a power law dependence for $\epsilon_N$,
one obtains
\begin{equation}
 \epsilon_N\propto N^{\gamma'}.\label{eq:eps-gammap}
\end{equation}
Hence the meaning of the parameter $\gamma^\prime$ follows, it is
the exponent of the leading finite-size correction to the
limit distribution. Clearly, $\gamma'<0$ is needed for 
power-like decay whereas a correction faster or slower
than power-law should be
characterized within the RG theory of linear perturbations by a $\gamma'=-\infty$ and $\gamma'=0$,
respectively.

%%%%%%%%%%%%%%%%%%%%%%%%%%%%%%%%%%%%%%%%%%%%%%%%%%%%%%%%%%%%%%%%%%%%%%%%%%%
\subsection{Marginal case near the FTG fixed point }
\label{sS:iid-marg-sd}
%%%%%%%%%%%%%%%%%%%%%%%%%%%%%%%%%%%%%%%%%%%%%%%%%%%%%%%%%%%%%%%%%%%%%%%%%%%

One expects that slower than power such as logarithmic corrections are also consistent with \eqref{eq:eps-pn} and they emerge in the  $\gamma'\to 0$ limit.
We study this case by considering the marginal situation near the FTG fixed point, as described in Sec.\ \ref{S:iid-nlrg}.  Again we take an $\epsilon_N$ function, and get in one RG step $\epsilon'=\epsilon_{pN}$. Expanding the reciprocal of (\ref{eq:eps-nl}) yields
in leading order
\begin{equation}
 \frac{1}{\epsilon_{pN}} =  \frac{1}{\epsilon_N} + (s-2)\ln p.
\end{equation}
This is a functional equation for $1/\epsilon_N$, whose solution is the
logarithm function, yielding
\begin{equation}
 \epsilon_N \approx  \frac{1}{(s-2)\ln N}.\label{eq:eps-nlin}
\end{equation}
Thus in the case $\gamma=\gamma'=0$ the natural assumption of analyticity in
$\epsilon$ leads to the cubic next-to-leading correction, and entails
the logarithmic decay of the finite-size correction.  
This can be considered as
the generic stable manifold about the FTG fixed point and explains the typically slow, $1/\ln N$, decay of the corrections.

Note that the above results consistently describe stable perturbations. Indeed,
(\ref{eq:eps-nl}) gives an $\epsilon'$ closer to zero than $\epsilon$, if
$\epsilon$ has the same sign as $s-2$.  But this is just what we obtained in
(\ref{eq:eps-nlin}).

A more general case is when the nonlinear extension in the marginal situation is \eqref{eq:eps-xi}.  Translating that to the
finite-size correction we have in the end
\begin{align}
 \epsilon_N &\approx \frac{A}{(\ln N)^{\frac{1}{\xi-1}}},\\
A&= \frac{\text{sgn}(s)}{[|s|(\xi-1)]^{\frac{1}{\xi-1}}}.
\label{eq:eps-log-xi}
\end{align}
Hence RG theory accounts for faster-than-logarithmic decay in the case of
an invariant manifold nonanalytic in $\epsilon$.  

The conclusions drawn from the results of the RG study beyond 
the linear region
are quite revealing. The correction characterizes how fast the parent
distribution approaches the FTG fixed point.  A quite slow but typical
case is  when the parent is attracted by the RG
transformation to an invariant manifold analytic in both $x$ and the
perturbation parameter $\epsilon$, yielding a finite-size correction
logarithmically decaying in $N$. If  the $\epsilon$-dependence is
not analytic in next-to-leading order then a larger power of the log and so
faster decay is obtained.  These cases correspond to $\gamma'=0$ and polynomial
corrections in $x$, quadratic in leading and cubic in next-to-leading order.
Faster, essentially power-law correction  is obtained if the convergence is
characterized by $\gamma'<0$, when the correction function is no longer
polynomial. Finally, faster than power decay corresponds to the case
$\gamma'\to -\infty$, thus it falls beyond the validity of the present RG
theory.

%%%%%%%%%%%%%%%%%%%%%%%%%%%%%%%%%%%%%%%%%%%%%%%%%%%%%%%%%%%%%%%%%%%%%%%%%%%
\subsection{Connection to the parent distribution}
\label{sS:iid-lin-parent}
%%%%%%%%%%%%%%%%%%%%%%%%%%%%%%%%%%%%%%%%%%%%%%%%%%%%%%%%%%%%%%%%%%%%%%%%%%%

It remains to be seen how the perturbation parameter 
$\epsilon_N$ can be determined from the parent distribution.  
That is, we start out from the classic extreme
value problem of the distribution of the maximum of $N$ i.i.d.\ variables, each
characterized by the same parent distribution.   Let us write the integrated
parent distribution as
\begin{equation}
\mu(z)= e^{-e^{-g(z)}}, \label{eq:g-def}
\end{equation}
then the distribution of the maximum of $N$ variables is
\begin{equation}
M^{\max}_N(z)=\mu(z)^N = e^{-Ne^{-g(z)}}.
\end{equation}
Let us assume that $\mu(z)$ belongs to the domain of attraction of the fixed
point  distribution $M(x;\gamma)$ from \eqref{eq:gevd}.  Then,
by using the linear transformation ensuring the standardization
\eqref{eq:mp-norm}
\begin{align}
z&= a_N x + b_N, \label{eq:lintrans}\\
g(b_N)&=\ln N,\label{eq:bn}\\
g'(b_N)&= 1/a_N\label{eq:an},
\end{align}
one obtains
\begin{equation}
 M^{\max}_N(a_Nx+b_N)=M_N(x) \to M(x;\gamma).
\end{equation}
Our basic assumption is that for large but finite $N$ the approach to the fixed
point goes along some eigendirection as
\begin{equation}
  M_N(x)\approx
M(x+\epsilon_N\psi(x;\gamma,\gamma');\gamma).\label{eq:final-appr}
\end{equation}
Here $\epsilon_N$ gives the order of the correction and is assumed to vanish for
large  $N$ as characterized by the exponent $\gamma'$.

Taking the double negative logarithm of both sides and using
\eqref{eq:fp-fab2} we get
\begin{equation}
\begin{split}
 g(a_Nx+b_N)-g(b_N) &\approx f(x;\gamma)\\&\,\, +\epsilon_N f'(x;\gamma)
\psi(x;\gamma,\gamma').
\end{split}
\label{eq:g-to-f-pert}
\end{equation}
In the next step we shall use the properties (\ref{eq:f-norm}) and
(\ref{eq:psi}) together with the equalities $f''(0;\gamma)=-\gamma$ and
(\ref{eq:an}).  Then double differentiation of \eqref{eq:g-to-f-pert} at
$x=0$ yields the perturbation parameter
\begin{equation}
 \epsilon_N = \gamma_N - \gamma  ,\label{eq:epsilonN}
\end{equation}
where
\begin{equation}
 \gamma_N = -\frac{g''(b_N)}{g'^2(b_N)}  = \frac{da_N}{db_N}. \label{eq:gammaN}
\end{equation}
The second equality follows from \eqref{eq:an} and is useful in practical
calculations.

To summarize the practical recipe, given a parent distribution,
one calculates $\gamma_N$ and if
\begin{equation}
  \gamma_N \to \gamma
\end{equation}
then the parent belongs to the domain of attraction of the fixed point with
parameter $\gamma$.  Then one determines $\epsilon_N$ from (\ref{eq:epsilonN})
at least to leading order, whose approach to zero is characterized by the
exponent $\gamma'$, see \eqref{eq:eps-gammap}.  This exponent also determines
the eigenfunction $\psi(x;\gamma,\gamma')$ in the final approach to the fixed
point, as it appears in \eqref{eq:final-appr}.  In the end, by linear
expansion  of \eqref{eq:final-appr}, we obtain the leading correction
to the fixed point distribution in the form
\begin{equation}
\begin{split}
  M_N(x)&\approx M(x;\gamma) + \epsilon_N  M_{1}(x;\gamma,\gamma') \\ &=
M(x;\gamma) +\epsilon_N P(x;\gamma)
\psi(x;\gamma,\gamma'),\label{eq:final-appr-add}
\end{split}
\end{equation}
corresponding to the density
\begin{equation}
\begin{split}
P_N(x)&= \frac{d}{dx}M_N(x)\approx P(x;\gamma) + \epsilon_N
P_{1}(x;\gamma,\gamma'). \label{eq:final-appr-add-dens}
\end{split}
\end{equation}
We relegate to a later work a more detailed discussion of 
the connection to the parent distributions.
Some examples of specific parents will be considered below.

The picture that emerges from the RG study complements the long-known
connection between the Fisher-Tippett max-stability condition and the limit
distributions.  By reformulating the former as the fixed point relation of an
appropriately defined RG transformation, we can then solve the eigenvalue
problem for distributions close to a fixed point. These eigenfunctions will
then appear as correction functions in the final approach to the limit
distribution for large batch sizes $N$.  One new parameter labeling the
eigenfunctions also emerged, the characteristic exponent of the convergence in
$N$.

The case of marginality, $\gamma'=0$, deserves a special mention.  
As it was discussed in Sec.\ \ref{S:iid-lin-mg}, nearby points 
on the fixed line are connected by marginal eigenfunctions.  
Using  \eqref{eq:evd-marginal} and substituting 
there $\epsilon$ by $\epsilon_N=\gamma_N-\gamma$ 
\eqref{eq:epsilonN}, we realize that the 
distribution \eqref{eq:final-appr-add} is, up to leading order, 
a fixed point distribution with a modified index
\begin{equation}  
M_N(x) \approx M(x;\gamma_N).   
\end{equation}  
This property was noted already by Fisher and Tippett \cite{FisherTippett:1928} in the case of the Gaussian parent, 
and studied since then in the mathematical literature, 
see \cite{Kaufmann:2000}. In the RG picture  such an asymptote 
means that the final approach to a fixed point is tangential 
to the fixed line, which happens when the finite-size correction 
decays slower than a power law.

%%%%%%%%%%%%%%%%%%%%%%%%%%%%%%%%%%%%%%%%%%%%%%%%%%%%%%%%%%%%%%%%%%%%%%%%%%%%
\subsection{Example for a parent in the FTG class: generalized exponential
asymptote}
\label{S:iid-examples-ftg-genexp}
%%%%%%%%%%%%%%%%%%%%%%%%%%%%%%%%%%%%%%%%%%%%%%%%%%%%%%%%%%%%%%%%%%%%%%%%%%%%

We demonstrate the asymptotic analysis above on the example of a parent
distribution, whose asymptote for $z\to\infty$ is of the following generalized exponential form
\begin{align}
 \mu(z) &\approx 1- B e^{-Az^{\delta}}/z^\theta,\ \ \
\delta>0,\label{eq:gen-exp}\\
 g(z)&=-\ln \left( -\ln \mu(z)\right) \approx A z^\delta + \theta\ln z-\ln B.
\label{eq:gen-exp-parent}
\end{align}
The leading singular parts of the shift and scale 
parameters are obtained from (\ref{eq:bn},\ref{eq:an}),
respectively, as
\begin{align}
b_N&=g^{-1}(\ln N)\approx\left( \frac{\ln N}{A}\right)^{1/\delta},\\
a_N &= 1/g'(b_N) \approx \frac{b_N^{1-\delta} }{A\delta} - \frac{\theta
b_N^{1-2\delta}}{A^2\delta^2} .
\end{align}
Next we calculate from \eqref{eq:gammaN} the effective parameter
\begin{equation}
\gamma_N = \frac{da_N}{db_N} \approx \frac{1-\delta}{A\delta}b_N^{-\delta}+ \frac{2\delta -1}{A^2\delta^2}\,\theta\, b_N^{-2\delta} .
\end{equation}
The limit gives of course
\begin{equation}
 \gamma_N \to \gamma=0,
\end{equation}
the parameter for the FTG fixed point. Hence the perturbation parameter 
to leading order is
\renewcommand{\arraystretch}{2.5}
\begin{align}
 \epsilon_N=\gamma_N \approx\left\lbrace
\begin{array}{lr}
\dfrac{1-\delta}{\delta\ln N}, &\delta\neq 1 \, , \\
\theta \dfrac{1}{\ln^2N}, & \delta=1 \, .\\
\end{array}\right.
\label{eq:eps-ftg}
\end{align}
We call the attention to the fact that the above expressions do not
depend on $A$ and $B$, they would enter only in the
next order in $1/\ln N$. 
It should also be noted that the case $\delta=1,\,\theta=0$ is special 
since it is not sufficiently specified by 
\eqref{eq:gen-exp} and further corrections to the asymptote need to
be included. A representative case is the pure exponential parent 
distribution which will be treated below in this section.

In the cases described by \eqref{eq:eps-ftg}, the decay is slower than
power law, thus
\begin{equation}
 \gamma' = 0.
\end{equation}
We can see that  the reciprocal logarithmic finite-size correction is valid for
all $\delta\neq 1$, so it is indeed typical, 
as surmised from nonlinear RG theory.
Note that, for $\delta\neq 1$, the value of $\theta$ does not enter the 
leading term in $\epsilon_N$, as can be seen from \eqref{eq:eps-ftg} and 
anticipated from the fact that the power factor $z^\theta$ 
in \eqref{eq:gen-exp} represents a next to leading 
correction in the exponent. Exceptionally, when $\delta=1$ the $\theta$ dependence does appear in the leading term of $\epsilon_N$ as shown in 
\eqref{eq:eps-ftg}.

A further result of the nonlinear RG was the EVS 
distribution to second order in
$\epsilon$, (\ref{eq:meps-nl}), with  $\psi_2(x;s)=sx^3/6$ as given by
(\ref{eq:psi-cub}).  Here the  coefficient $s$ is related to the magnitude of
the perturbation parameter $\epsilon_N$ by Eq.\  (\ref{eq:eps-nlin}).  From the
comparison of Eqs.\ (\ref{eq:eps-nlin},\ref{eq:eps-ftg}) we get
\begin{equation}
s=\frac{\delta-2}{\delta-1}.
\end{equation}
The parameter $s$ defines the second order finite-size correction for the
extreme value distribution. Note that, for consistency,
an expansion up order $1/\ln^2\!\!N$ should also involve  
the expansion of $\epsilon_N$ to this order.

The case $\delta=1$ with $\theta\neq 0$ corresponds to a power-corrected
exponential asymptote of the parent, when the perturbation parameter
$\epsilon_N$ exhibits reciprocal square logarithmic  convergence.  This is the
situation treated in the end of Sec.\ \ref{S:iid-nlrg}.  Comparison of
(\ref{eq:eps-log-xi}) and the second line of (\ref{eq:eps-ftg}) gives
\begin{eqnarray}
\xi &=& \frac 32,\\
 s&=&-\frac{2\text{sgn}(\theta)}{\sqrt{|\theta|}}.
\end{eqnarray}

The family of asymptotes for the parent distribution (\ref{eq:gen-exp})
comprises some notable cases.  First we mention the Gaussian parent having
$\delta=2$, $\theta=1$, thus  by (\ref{eq:eps-ftg}) we get $\epsilon_N\approx -
1/2\ln N$ and so the linear correction defined in
\eqref{eq:final-appr-add} becomes
\begin{equation}
 \epsilon_N M_1(x) \approx \frac{x^2}{4\ln N}e^{-x-e^{-x}}.
\label{eq:gauss-unscaled}
\end{equation}
The same result is obtained for the Rayleigh parent distribution,
$\mu(z)=1-e^{-x^2}$, where $\delta=2$, $\theta=0$. Another situation, which was
used in \cite{GyorgyiETAL:2008}, is $\delta=1$, $\theta=1$, describing the
exponential distribution with a $1/z$ factor.  This is in fact the asymptote of
the distribution of the sizes of connected components in subcritical site
percolation.  In this case (\ref{eq:eps-ftg}) yields $\epsilon_N\approx
1/\ln^2\!\! N$ and the correction to the distribution becomes
\begin{equation}
 \epsilon_N M_1(x) \approx - \frac{x^2}{2\ln^2\!\! N}e^{-x-e^{-x}}.
\label{eq:gauss-unscaled-2}
\end{equation}
Finally we mention the pure exponential decay, corresponding to $\delta=1$ and
$\theta=0$.  In the above approximation $\epsilon_N=0$, thus one has to
go to a higher order in $g(z)$.  As a notable example we consider the
exponential parent $\mu(z)=1-e^{-Az}$, where $g(z)\approx Az -
\frac{1}{2}e^{-Az}$, and so $b_N\approx A^{-1}\ln N$ and $a_N\approx  A^{-1} -
(2A)^{-1}e^{-Az}$.  Then the effective $\gamma_N=da_N/db_N \approx (2N)^{-1}$,
which equals now $\epsilon_N$. Thus the decay index is $\gamma'=-1$ and
\begin{equation}
 \epsilon_N M_{1}(x) \approx  \frac{1-x-e^{-x}}{2N} e^{-x-e^{-x}}
\label{eq:exp-unscaled}
\end{equation}
gives the leading correction to the asymptotic FTG fixed point distribution.
Note that the above correction functions are understood with the standardization
\eqref{eq:stand}, so in empirical data analysis a further linear change of
variable may have to be considered.

\begin{figure}[htb]
\includegraphics[width=8.2cm,clip]{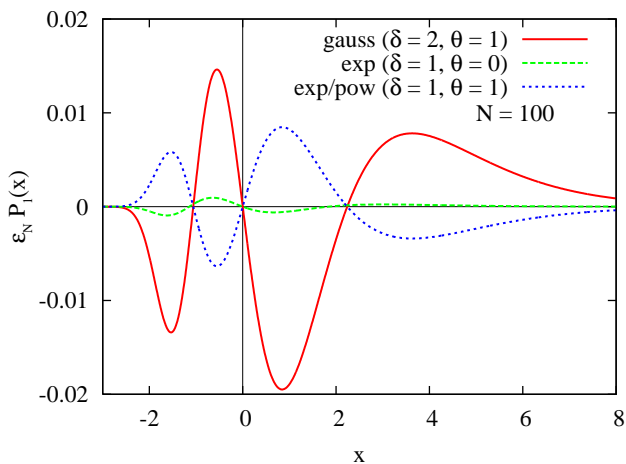}
\caption{(Color online) First order corrections 
$\epsilon_N P_1(x,\gamma,\gamma')=[\epsilon_NM_1(x,\gamma,\gamma')]'$
to the limit distribution density in the FTG class, $\gamma =0$, for 
the generalized exponential case
$\mu(z) \approx 1- B e^{-Az^{\delta}}/z^\theta$.
The different magnitudes
of the shape corrections are illustrated by evaluating 
the amplitude $\epsilon_N$ at the same batch-size $N=100$. Note 
that while the curves marked as {\it gauss} and {\it exp} have the 
same shape with different prefactors, the functional form of {\it exp/pow}
is slightly different. 
\label{F:ftg-exp}}
\end{figure}

Fig.\ref{F:ftg-exp} displays the corrections to limit 
distributions for the three examples treated above (\ref{eq:gauss-unscaled},\ref{eq:gauss-unscaled-2},\ref{eq:exp-unscaled})
using the standardization \eqref{eq:stand}
adopted here for theoretical reasons. The aim here is to illustrate 
empirical plots of the first order corrections including
the amplitude $\epsilon_N$.
One can clearly see that the fastest convergence 
occurs for the exponential parent distribution. Furthermore,
while the functional forms for
the Gaussian ($\delta =2, \,
\theta=1$) and for the power law corrected exponential ($\delta =1, \,
\theta=1$) are the same, the difference in the sign of the correction
makes the empirical distinction apparent.

%%%%%%%%%%%%%%%%%%%%%%%%%%%%%%%%%%%%%%%%%%%%%%%%%%%%%%%%%%%%%%%%%%%%%%%%%%%%
\section{Unstable eigenfunctions and invariant manifolds}
\label{S:iid-unst}
%%%%%%%%%%%%%%%%%%%%%%%%%%%%%%%%%%%%%%%%%%%%%%%%%%%%%%%%%%%%%%%%%%%%%%%%%%%%

So far we have been concentrating on the small neighborhood of
the fixed line. 
However, regions far away from the critical line where the space of
distribution functions with respect to the RG transformation
becomes more complex, are also of interest. 
In order to explore those regions, we should select special
parent distributions. Upon the action of the RG transformation with
$p>1$, the parent distributions we considered
approached a fixed point along a stable eigendirection.
This is the last surviving  perturbation and
represents the finite-size correction to the limit
distribution for large batch size $N$, and is characterized
by $\gamma'\le 0$. On the other hand, eigenvalues
$\lambda=p^{\gamma'}>1$ also exist
corresponding to  $\gamma'>0$.  So the question naturally arises:
what happens, if we start out from a parent distribution
that is close to a fixed point
labeled by $\gamma$, and differs from it in an unstable
eigenfunction $\psi(x;{\gamma,\gamma'})$ with
a small magnitude $\epsilon$.
Initially, the distribution  will move away from the
fixed point by the action of the linearized RG but
the evolution becomes nontrivial at later stages
due to the nonlinearities present.
One can nevertheless pose the question
whether parent distribution belongs to the domain of attraction
of a standard EVS fixed point.

By evolving a distribution containing an unstable eigendirection as a
perturbation about a fixed point, an invariant unstable manifold emerges. 
This manifold can be determined order by order in the perturbation 
parameter $\epsilon$, by the successive application of the RG 
transformation.  An ambiguity emerges, however, as to how a 
perturbation proportional to $\epsilon$ is
introduced. This is exemplified by two natural choices of  distributions
(\ref{eq:meps},\ref{eq:peps}), redisplayed with an eigenfunction as
perturbation
\begin{subequations}
\begin{align}
\mu_1(z)&= M(z+\epsilon\psi(z;\gamma,\gamma');\gamma), \label{eq:unst-parent1}
\\
\mu_2(z)&= M(z;\gamma)+\epsilon P(z;\gamma)\psi(z;\gamma,\gamma'),
\label{eq:unst-parent2}
\end{align}\label{eq:unst-parent}
\end{subequations}
\!\!where $M$ is a fixed point distribution, $P$ its density, $\psi$ is given by
\eqref{eq:psi-rg}, and $\epsilon$ is a small, but fixed, parameter.  While
these distributions are the same up to linear order in $\epsilon$, they 
generically differ
in higher order terms and in their large-$z$ asymptote.  
Thus it matters which distribution we start
with, because they will track different manifolds. More variants equivalent to linear order are also possible, but we restrict our attention to the
above two types.

A problem to be considered when various linear perturbations
are introduced, as
in Eqs.\ \eqref{eq:unst-parent}, is that we should have a
valid (monotonic from
0 to 1)  distribution for finite $\epsilon$ in some support in $z$. Since the
eigenfunctions $\psi(z;\gamma ,\gamma')$ 
are non-monotonic in $z$, extra care should be exercised. 
In particular, the equivalence of
the two forms in \eqref{eq:unst-parent} holds only for 
$|\epsilon\psi(z;\gamma ,\gamma')|\ll |z|$,
but this condition is generically not fulfilled in the entire support. 
We circumvent the above problem by considering particular
parent distributions, specifying which form in 
\eqref{eq:unst-parent} should be
understood, and determining some allowed sets of the parameters
$\epsilon,\gamma'$.  Below, we shall discuss separately  the unstable
invariant manifolds emanating from the three classes of fixed point
distributions.

%%%%%%%%%%%%%%%%%%%%%%%%%%%%%%%%%%%%%%%%%%%%%%%%%%%%%%%%%%%%%%%%%%%%%%%%%%%%
\subsection{Instability near the FTG fixed point}
\label{S:iid-unst-ftg}
%%%%%%%%%%%%%%%%%%%%%%%%%%%%%%%%%%%%%%%%%%%%%%%%%%%%%%%%%%%%%%%%%%%%%%%%%%%%

First we shall consider the parent of the form \eqref{eq:unst-parent1}
\begin{equation}
\mu(z)=\mu_1(z)= e^{-e^{-z-\epsilon\psi(z)}}, \label{eq:unst-ftg-parent}
\end{equation}
where $\epsilon$ is assumed small but finite and fixed, and
\begin{equation}
\psi(z)=\psi(z;0,\gamma')=\frac{1}{\gamma'^2} \left( 1+\gamma'z-e^{\gamma'z}\right)
\label{eq:unst-ftg-psi}
\end{equation}
is an unstable eigenfunction with $\gamma'>0$.  It is easy to see that for
$0<-\epsilon<\gamma'$ we have a strictly increasing $\mu(z)$ for all $z$, 
so the support extends over the entire real axis.  
For sufficiently large $z$, however, in
the region of importance for extreme statistics, one has $\epsilon\psi(z) \gg
z$, so an expansion in $\epsilon$ is no longer legitimate and
\eqref{eq:unst-ftg-parent} is not equivalent to the
corresponding formula in \eqref{eq:unst-parent2} even for small $\epsilon$.

We now would like to determine whether the parent is attracted to a fixed point,
and, if so, what its finite-size correction is.  The associated parameters will be
denoted by $\tilde\gamma$, labeling the attracting fixed point, and
$\tilde\gamma'$, specifying the eigenfunction of the final approach.  These parameters can be determined by means of the relations given in Sec.\ \ref{sS:iid-lin-parent}, namely
\begin{subequations}
\begin{align}
g(z) &= -\ln (-\ln \mu(z) )\nonumber \\
&= z + \epsilon\psi(z)  \approx \frac{ |\epsilon|}{\gamma'^2}
e^{\gamma'z},\label{eq:gz-ftg}\\
b_N&= g^{-1}(\ln N) \approx  \frac{1}{\gamma'}  
\ln{ \left( \frac{\gamma'^2}{|\epsilon|} \ln N \right)}, \label{eq:b-ftg}\\
a_N&= \frac {1}{g'(b_N)}\approx \frac{\gamma'^2}{|\epsilon|} e^{-\gamma'b_N}.
\label{eq:a-ftg}
\end{align}
\end{subequations}
Convergence is characterized by the effective $\gamma_N$  defined in
\eqref{eq:gammaN}, which now approaches $\tilde\gamma$, and by the
perturbation parameter $\epsilon_N$ given in  \eqref{eq:epsilonN}.  We
now have
\begin{align}
\gamma_N=\frac{da_N}{db_N} = \tilde\gamma + \epsilon_N \approx
-\frac{\gamma'^2}{|\epsilon|} e^{-\gamma'b_N}.
\end{align}
Hence we can easily determine the indices of universality
\begin{subequations}\begin{align}
 \gamma_N &\to \tilde\gamma =0,\\
 \epsilon_N &= \gamma_N \approx -\frac{1}{\ln N}, \label{eq:um-ftg-gN}\\
 \tilde\gamma' &=0. \label{eq:um-ftg-gammap1}
\end{align}
 \end{subequations}
The third line follows from the slower than power decay in the second one.
Note that the function $g(z)$ discussed in this subsection 
diverges faster than any power and, accordingly, the result
\eqref{eq:um-ftg-gN} can be obtained as the $\delta\to\infty$ 
limit of \eqref{eq:eps-ftg}.

\begin{figure}[htb]
\includegraphics[width=8.3cm]{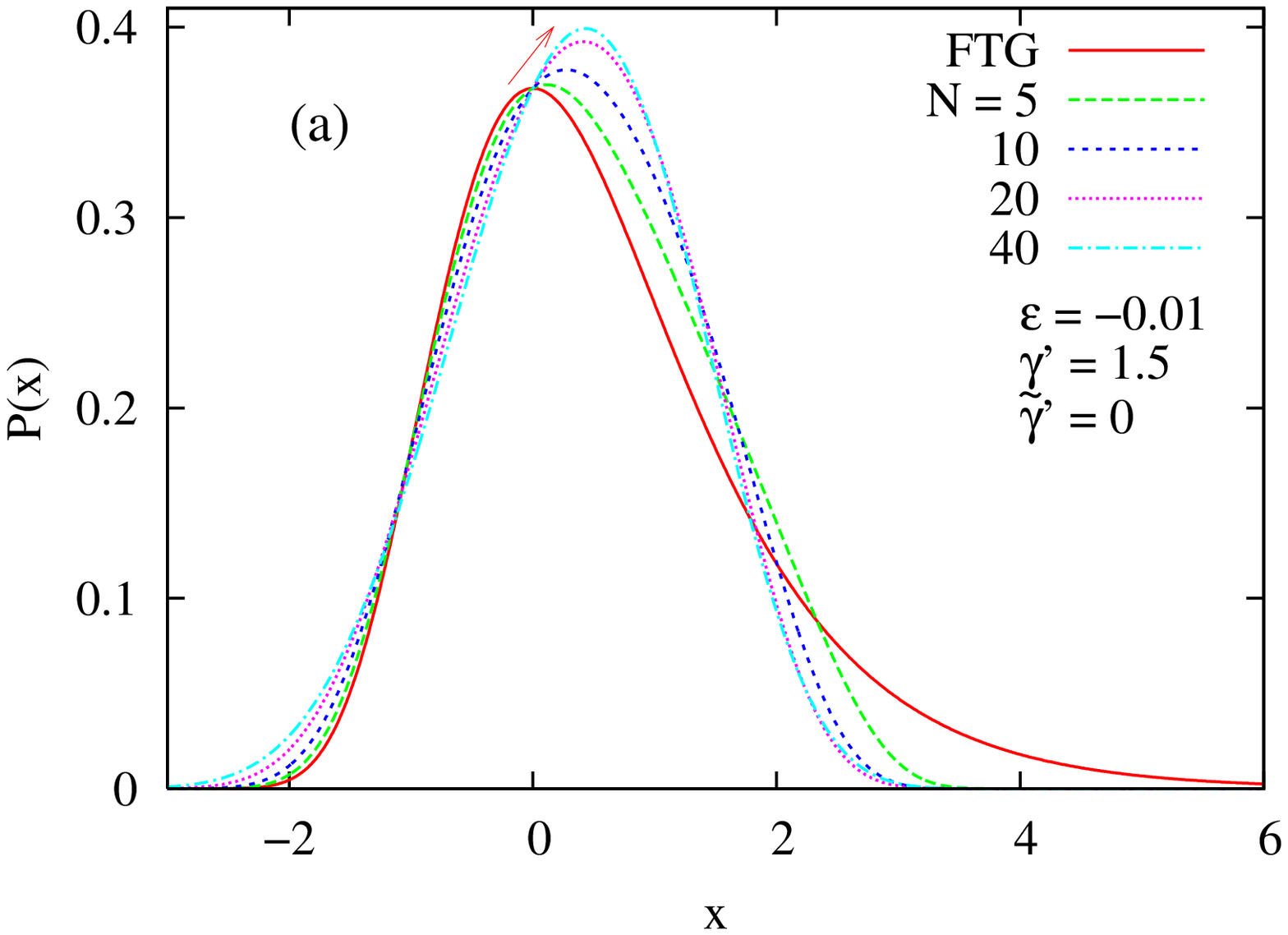}
\includegraphics[width=8.3cm]{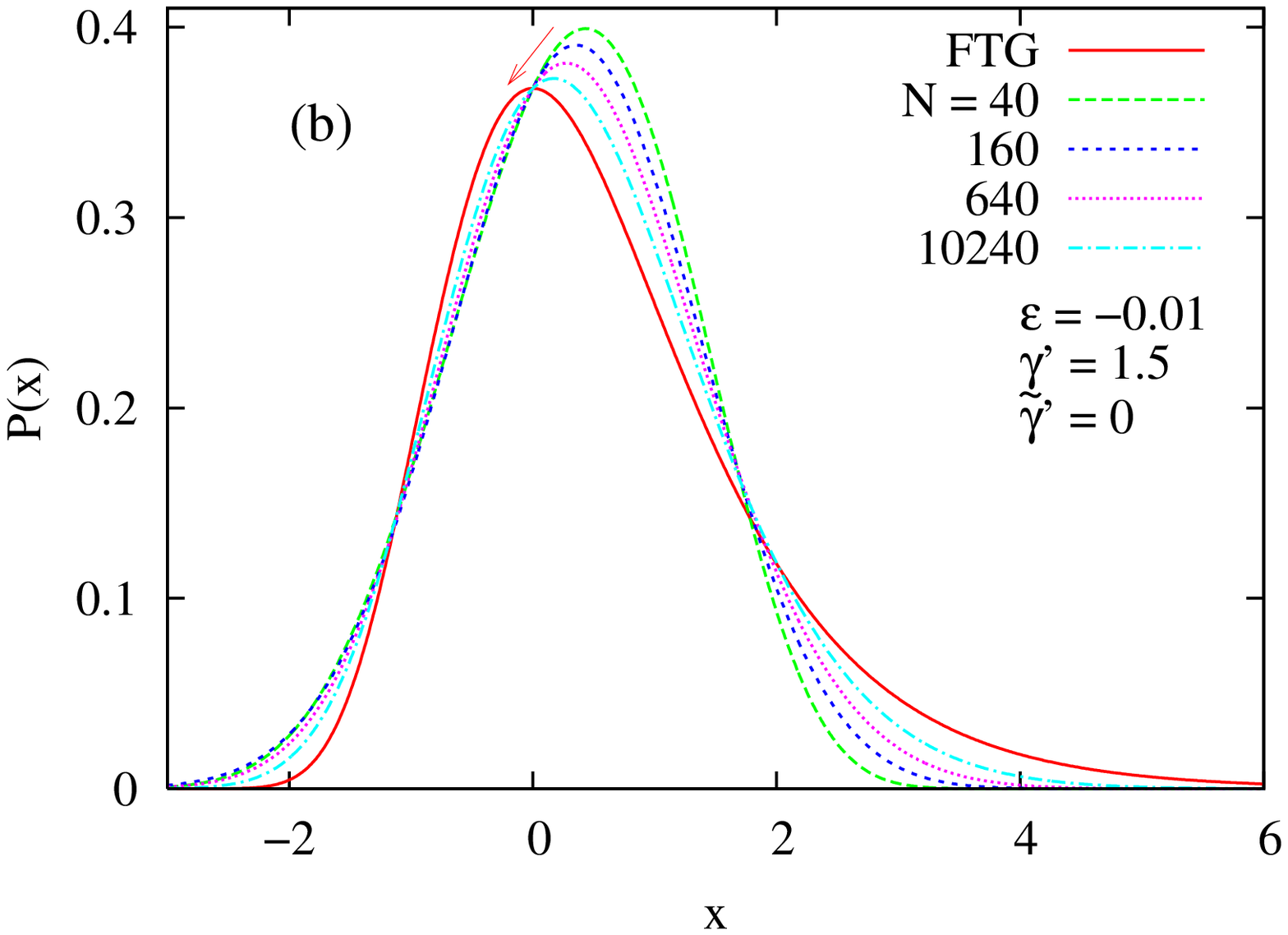}
\caption{(Color online) Probability densities on the 
invariant manifold starting from the
unstable parent \eqref{eq:unst-ftg-parent} with 
\eqref{eq:unst-ftg-psi}:  (a) A sequence of densities
diverging from FTG. (b)  A sequence of densities returning towards FTG.
The arrows mark the direction of $N$ increasing.
The initial $\epsilon,\gamma'$ and final $\tilde\gamma'$ parameters are
indicated.  Since the effective $\gamma_N$
tends to zero logarithmically slowly, see \eqref{eq:um-ftg-gN}, one has to go
up to relatively high $N$ to demonstrate convergence to the 
FTG distribution. 
\label{F:ftg-um}}
\end{figure}

Summarizing, if we start out from the parent \eqref{eq:unst-ftg-parent} 
close to the
FTG fixed point, but differing from it in an unstable eigenfunction then the
RG transformation drives it to the same fixed point  along the
eigendirection with marginal eigenvalue $1$.  That is, this unstable
manifold with a $\gamma'>0$ loops back along  the marginally stable one with
$\tilde\gamma'=0$.  We illustrate this effect as follows.  On Fig.\
\ref{F:ftg-um}a the initial $\gamma'=1.5$ and $\epsilon=-0.01$ was taken,
and the extreme value density of $N$ variables displayed for some increasing
$N$, showing the increasing deviation from FTG.  
The $N$ here corresponds to the
parameter $p$ of the RG transformation \eqref{eq:def-rg}.   Figure
\ref{F:ftg-um}b, on the other hand, shows that for further increasing $N$
the sequence of distributions approach FTG, thus completing 
the loop of the invariant manifold.

\begin{figure}[htb]
\includegraphics[width=8.3cm]{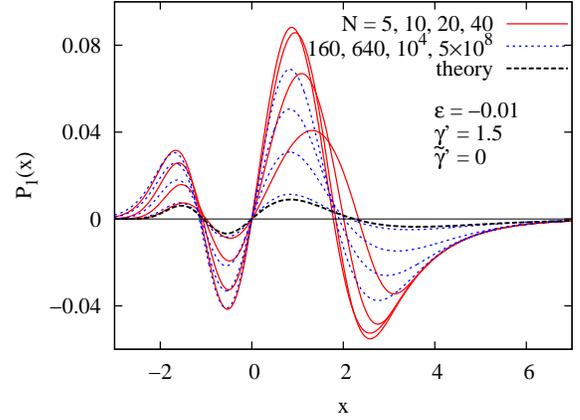}
\caption{(Color online) The differences from FTG of the 
densities plotted in Fig.~\ref{F:ftg-um}.  
The magnitudes of the solid red curves increase while those 
of the dotted blue ones decrease with $N$.  
It is apparent that the initial and final shape corrections have different
functional form (the solid blue and dotted red curves with smallest 
magnitude).  
For illustration, the marginal correction function (``theory'') 
with the amplitude
$\epsilon_N=-1/\ln N$ ($N=5\times 10^8$) 
from \eqref{eq:um-ftg-gN} is also displayed (dashed black line), well
approximating the last dotted blue curve.
\label{F:ftg-um-diff}}
\end{figure}

The loop in function space can be further visualized by 
plotting the evolution of difference from the FTG limit distribution 
as $N$ changes (Fig.\ref{F:ftg-um-diff}). A feature which can be easily 
seen on this plot is that the initial and the final small deviations
from the FTG have different functional forms corresponding to the 
different values of $\gamma'$ and $\tilde \gamma'$. 

\begin{figure}[htb]
\includegraphics[width=8.3cm]{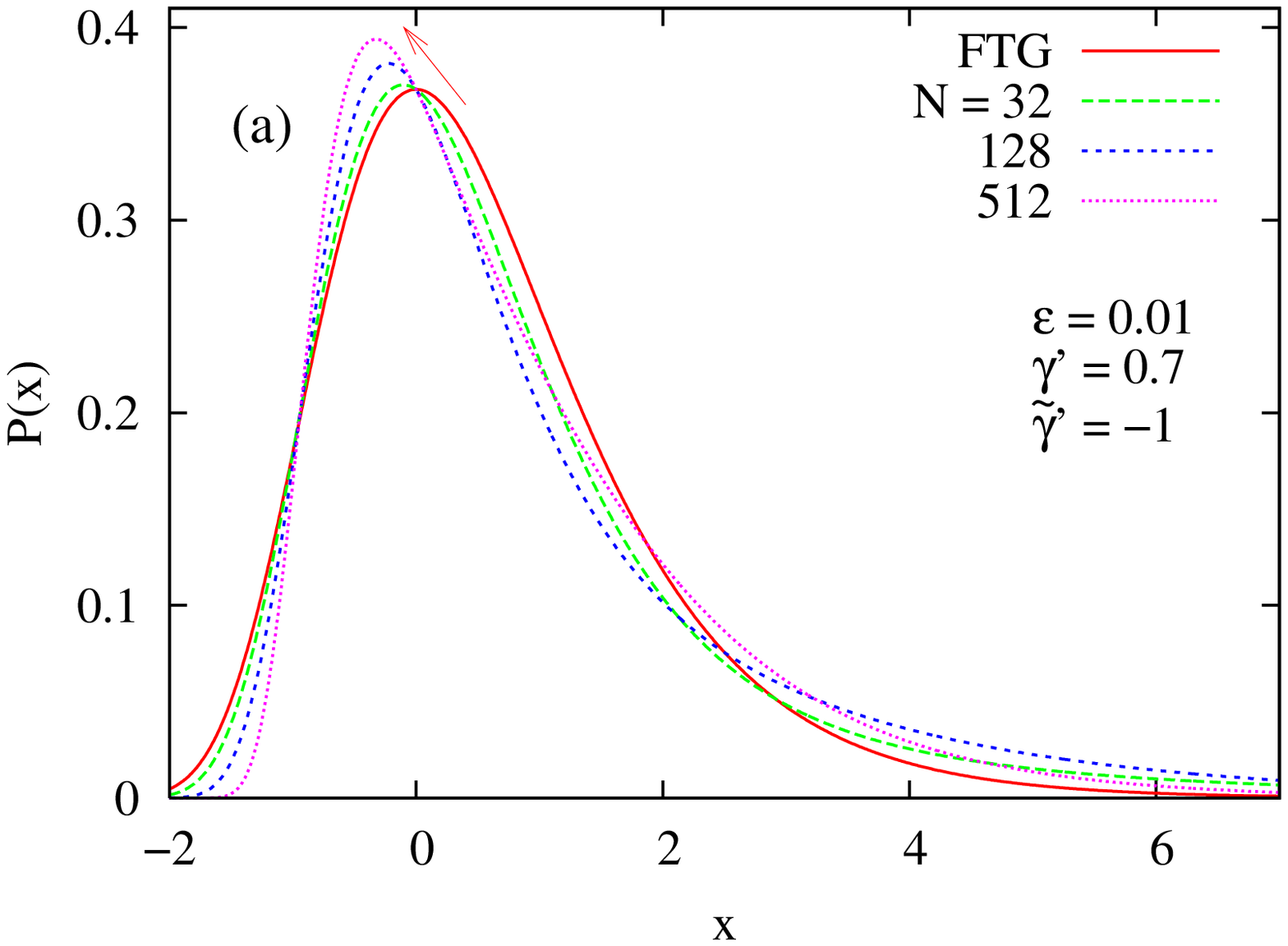}
\includegraphics[width=8.3cm]{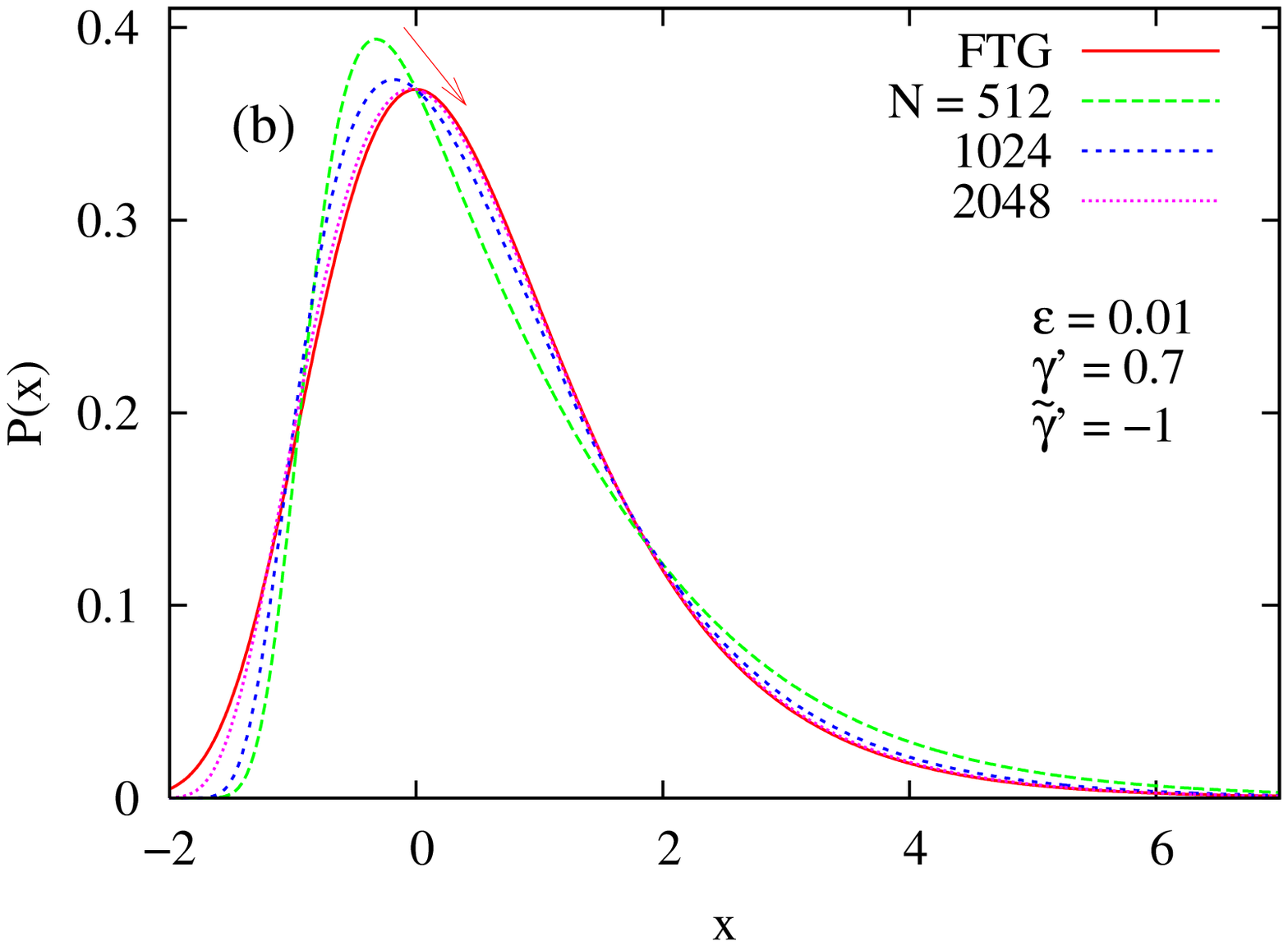}
\caption{(Color online) Probability densities on the 
invariant manifold starting from the
unstable parent \eqref{eq:unst-ftg-parent2} with 
\eqref{eq:unst-ftg-psi}: (a) A sequence of densities
diverging from FTG. (b) A sequence of densities returning
towards FTG.  Arrows mark the direction of increasing $N$. The convergence
to FTG is markedly faster than that along the marginal
eigendirection in Fig.~\ref{F:ftg-um}.
\label{F:ftg-um2}}
\end{figure}

In order to demonstrate the difference of the forms \eqref{eq:unst-parent1} and \eqref{eq:unst-parent2}, next we consider the same perturbation as in \eqref{eq:unst-ftg-parent} but now defined through the form \eqref{eq:unst-parent2}. That is, we take the parent
\begin{equation}
\mu(z)=\mu_2(z)= e^{-e^{-z}}\left[ 1 + \epsilon \psi(z) e^{-z}\right] \,,
\label{eq:unst-ftg-parent2}
\end{equation}
where \eqref{eq:unst-ftg-psi} is implied.
To have $\mu(z)\to 1$ from below for $z\to \infty$, we should set $1>\gamma'>0$ and $\epsilon> 0$.  Note that formula \eqref{eq:unst-ftg-parent2} would yield negative $\mu(z)$ for large negative arguments, but $\mu(z)$ becomes positive for smaller negative
arguments, and a  sufficiently small $\epsilon$ ensures that thenceforth
$\mu(z)$ is strictly increasing.  Thus if we discard the region of negativity
from the support then the function so obtained becomes a legitimate integrated
probability distribution.   For small enough $\epsilon$ it is obviously an
unstable direction, and thus the source of an invariant manifold after RG
transformations.  Its large $z$ behavior  can be characterized by the function
$g(z)$ in \eqref{eq:g-def} as
\begin{equation}
\begin{split}
g(z) &= -\ln (-\ln \mu(z) )\\ &\approx (1-\gamma') z + \ln
\frac{\gamma'^2}{\epsilon} + \gamma' z e^{-\gamma'z} \\ & \ \ \ -
\frac{\epsilon}{2\gamma'^2} e^{-(1-\gamma')z}, \label{eq:unst-ftg-g2}
 \end{split}
\end{equation}
where we kept competing exponential terms.  Hence, according to the procedure
set forth in Sec.\ \ref{sS:iid-lin-parent}, we obtain
\begin{equation}
 b_N= g^{-1}(\ln N) \approx  \frac{\ln \left( \epsilon N/\gamma'^2\right)
}{1-\gamma'}, \label{eq:b2-ftg}
\end{equation}
and
\renewcommand{\arraystretch}{2}
\begin{equation}
 \begin{split}
 \gamma_N&= - \frac{g''(b_N)}{g'^2(b_N)} \\
 &\approx \left\{ \begin{array}{lr}
 c\, N^{-\frac{\gamma'}{1-\gamma'}} \ln N , \ \ &
0 < \gamma' \leq 1/2,\\
 \dfrac{1}{2N} , \ \  & 1/2 < \gamma' < 1,
 \end{array}
 \right.
\label{eq:gammaN2-ftg}
 \end{split}
\end{equation}
where the constant $c$ depends on $\epsilon$ and $\gamma'$.
Since $\gamma_N\to \tilde\gamma=0$, again the FTG fixed
point is approached.  Thus we have $\epsilon_N=\gamma_N$, whose
decay is characterized by the index
\begin{equation}
 \tilde \gamma'= \left\{ \begin{array}{lr}
 -\dfrac{\gamma'}{1-\gamma'}, \ \ & 0 < \gamma' \leq 1/2,\\
 -1, \ \  & 1/2 \leq \gamma' < 1.
 \end{array}
 \right.
 \label{eq:gammap-ftg2}
\end{equation}
Hence a nonmarginal asymptotic index, $\tilde\gamma'<0$, is obtained, which
characterizes the last surviving eigenfunction.  This should be contrasted with
the previously discussed case of \eqref{eq:unst-ftg-parent}, where independently
of the initial $\gamma'$ we had $\tilde\gamma'=0$ as shown in
\eqref{eq:um-ftg-gammap1}. That demonstrates the
inequivalence of the distributions with unstable perturbations,
\eqref{eq:unst-parent1} and \eqref{eq:unst-parent2} in the sense that depending
 on whether we apply the perturbation additively or in the argument, different
limiting properties emerge.  The unstable manifold generated from
\eqref{eq:unst-ftg-parent2} is illustrated on Figs.\
\ref{F:ftg-um2}.

%%%%%%%%%%%%%%%%%%%%%%%%%%%%%%%%%%%%%%%%%%%%%%%%%%%%%%%%%%%%%%%%%%%%%%%%%%%%
\subsection{Instability near the FTF fixed point}
\label{S:iid-unst-ftf}
%%%%%%%%%%%%%%%%%%%%%%%%%%%%%%%%%%%%%%%%%%%%%%%%%%%%%%%%%%%%%%%%%%%%%%%%%%%%

Consider the parent of the type \eqref{eq:unst-parent1}
\begin{equation}
\mu(z)= \mu_1(z)=e^{-(1+\gamma
z+\epsilon\gamma\psi(z))^{-1/\gamma}},\label{eq:mu-unstable-ftf}
\end{equation}
where $\psi(z)$ abbreviates $\psi(z;\gamma ,\gamma')$ as given by (\ref{eq:psi-rg}) and $\gamma,\gamma'>0$. This is a
legitimate integrated distribution in $[-1/\gamma,\infty)$ if
$\gamma'>-\epsilon>0$.  Note that  $\mu(z)$ has a positive value 
at the lower border $-1/\gamma$, below it we define $\mu(z)$ 
to be zero.

\begin{figure}[htb]
\includegraphics[width=8.3cm]{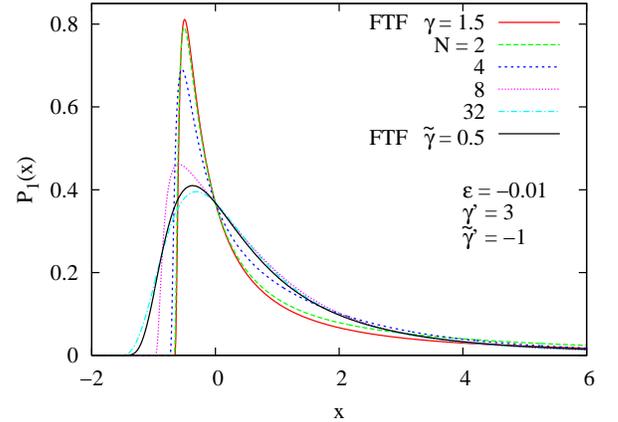}
\caption{(Color online) Sequence of densities on an 
invariant manifold of the form
\eqref{eq:mu-unstable-ftf}, with increasing $N$ tracking the RG
trajectory.  In variance with the FTG case, the starting
and final distributions are different here.  
Note that while convergence to the latter is well
demonstrated, the
maximum decreases to below the limit before the final approach.
This is related to the fact that the effective
$\gamma_N$ in \eqref{eq:um-ftf-gammaN1} approaches the limit from below.
\label{F:ftf-um}}
\end{figure}

We again denote by $\tilde\gamma,\tilde\gamma'$ the characteristic parameters
for the above parent. The procedure for calculating them was described in Sec.\ \ref{sS:iid-lin-parent}.    We now have
\begin{align}
g(z) &= -\ln (-\ln \mu(z) )\nonumber \\ &=\frac 1\gamma \ln(1+\gamma z+ \gamma\epsilon\psi(z)),\label{eq:gz-ftf} \\
e^{\gamma g(b_N)} & =  1+\gamma b_N + \gamma \epsilon \psi(b_N) = N^\gamma, \label{eq:um-ftf-b}\\
a_N&= \frac {1}{g'(b_N)} = \frac{1+\gamma b_N
+\gamma\epsilon\psi(b_N)}{1+\epsilon \psi'(b_N)}.
\end{align}
Expanding the above formula in $y_N=1+\gamma b_N$,
using \eqref{eq:gammaN} and
the expression for $\psi(z)$ in \eqref{eq:psi-rg}, we obtain
\begin{align}
\gamma_N&=\frac{da_N}{db_N} = \gamma \frac{da_N}{dy_N}
\nonumber \\
&\approx  \frac{\gamma^2}{\gamma'+\gamma} + \frac{\gamma-\gamma'}
{\gamma+\gamma'} \gamma'\left( \frac{\gamma'}{|\epsilon|} -
1\right) y_N^{-\gamma'/\gamma} \nonumber \\
 &+ O(y_N^{-\gamma'/\gamma-1}) + O(y_N^{-2\gamma'/\gamma}).
\label{eq:um-ftf-gammaN1}\end{align}
The large-$N$ limit determines the index of the attracting fixed point
\begin{equation}
 \tilde\gamma = \frac{\gamma^2}{\gamma'+\gamma}.\label{eq:gammatilde-ftf}
\end{equation}
This is just the asymptotic exponent in  $b_N\propto N^{\tilde\gamma}$, which
also can be obtained from \eqref{eq:um-ftf-b} and the dominant term in
\eqref{eq:psi-rg}. Next, we have to determine the
leading power of $N$ in
\begin{equation}
\epsilon_N = \gamma_N - \tilde \gamma \propto N^{\tilde\gamma'}.
\end{equation}
One should distinguish the case $\gamma'=\gamma$, the amplitude
of the term $y_N^{-\gamma'/\gamma}$ vanishes in
\eqref{eq:um-ftf-gammaN1}, and the leading correction
becomes $y_N^{-2}$.   Using $b_N\propto
y_N$ in  \eqref{eq:um-ftf-gammaN1} we finally obtain
\begin{equation}
 \tilde\gamma' = \left\{ \begin{array}{lr}
-\dfrac{\gamma\gamma'}{\gamma'+\gamma}, & \gamma'\neq\gamma, \\
-\gamma', & \gamma'=\gamma.
\end{array}
\right. \label{eq:gammaprimetilde-ftf}
\end{equation}
We illustrate the above described invariant manifold
on Fig.\ \ref{F:ftf-um}, with starting parameters
$\gamma=1.5,\gamma'=3$ resulting in final ones
$\tilde\gamma=0.5,\tilde\gamma'=-1$.

\begin{figure}[htb]
\includegraphics[width=8.3cm]{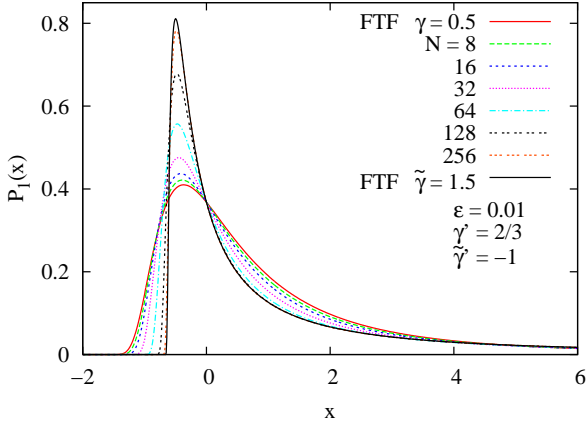}
\caption{(Color online) Sequence of densities on an 
invariant manifold starting out from
\eqref{eq:unst-ftf-parent2}, where increasing $N$ follows the RG trajectory.
Again, the starting and final distributions are different,
but in contrast with the case \eqref{eq:mu-unstable-ftf},
illustrated on Fig.\ \ref{F:ftf-um}, the
final $\tilde\gamma$ is larger than the initial $\gamma$.
\label{F:ftf-um2}}
\end{figure}

In order to demonstrate the inequivalence of \eqref{eq:unst-parent1} and
\eqref{eq:unst-parent2} also in the FTF case, we consider the example
\begin{equation}
\mu(z) = \mu_2(z) = e^{-(1+\gamma z)^{-{1}/{\gamma}}}
\left[ 1 + \epsilon \psi(z)
(1+\gamma z)^{-{1}/{\gamma} -1}\right]. \label{eq:unst-ftf-parent2}
\end{equation}
Given the form of $\psi$ in \eqref{eq:psi-rg},  the decay of the perturbation
for large $z$ requires $\gamma'<1$. It is easy to see that
near the lower border of the support of the unperturbed limit distribution, in
some interval for $z<0$, this function is negative.  Then for increasing
argument it becomes positive, and for not too large $\epsilon$ it remains
monotonic thereafter.  Since the EVS limit behavior depends on the large-$z$
asymptote, we simply discard the region of negativity from the support,
thus we wind up with a legitimate parent distribution, given by
\eqref{eq:unst-ftf-parent2} whenever it is non-negative. We
keep only the asymptotic terms necessary to determine the final
indices
\begin{equation}
\begin{split}
g(z) &= -\ln (-\ln \mu(z) ) \\ &\approx \frac{1-\gamma'}{\gamma}
\ln(1+\gamma z) + c_1 \\ &+ c_2(1+\gamma z)^{-\gamma'/\gamma} +c_3(1+\gamma
z)^{-(1-\gamma')/\gamma},
 \end{split} \label{eq:unst-ftf-g2}
\end{equation}
where the constants $c_j$ depend on the parameters $\epsilon, \gamma, \gamma'$.
Repeating the steps of
the procedure set forth in Sec.\ \ref{sS:iid-lin-parent}, we find
\begin{subequations}
 \begin{align}
\tilde\gamma &= \dfrac{\gamma}{1-\gamma'}, \\
 \tilde \gamma'&= \left\{ \begin{array}{lr}
 -\dfrac{\gamma'}{1-\gamma'}, \ \ & 0 \leq \gamma' \leq 1/2,\\
 -1, \ \  & 1/2 \leq \gamma' < 1.
 \end{array}
 \right.
\end{align}
\label{eq:gammatildes-ftf2}
\end{subequations}
On Fig.\ \ref{F:ftf-um2} we illustrate this case with
$\epsilon=0.01,\gamma=0.5,\gamma'=2/3$.  The asymptotic indices
can be read off
from Eq,\ \eqref{eq:gammatildes-ftf2} as
$\tilde\gamma=1.5$ and $\tilde\gamma'=-1$.

Summarizing,  the parent  with perturbed argument \eqref{eq:mu-unstable-ftf},
representing an unstable function near an FTF  fixed point with $\gamma>0$ and
having an instability parameter $\gamma'>0$, will eventually be attracted to
another  FTF limit distribution with $\tilde\gamma$ as in
(\ref{eq:gammatilde-ftf}).  The final approach
happens along the stable manifold with parameter $\tilde\gamma'$ as in
(\ref{eq:gammaprimetilde-ftf}).  In contrast, if we additively perturb the
fixed point and take for the parent
\eqref{eq:unst-ftf-parent2}, we get a different set of indices $\tilde\gamma,
\tilde\gamma'$, as given in (\ref{eq:gammatildes-ftf2}).

%%%%%%%%%%%%%%%%%%%%%%%%%%%%%%%%%%%%%%%%%%%%%%%%%%%%%%%%%%%%%%%%%%%%%%%%%%%%
\subsection{Instability near the FTW fixed point}
\label{S:iid-unst-ftw}
%%%%%%%%%%%%%%%%%%%%%%%%%%%%%%%%%%%%%%%%%%%%%%%%%%%%%%%%%%%%%%%%%%%%%%%%%%%%

In the FTW case, the two types of parent
distributions, (\ref{eq:unst-parent1})
and (\ref{eq:unst-parent2}) have the same formulas as for the FTF class,
\eqref{eq:mu-unstable-ftf} and \eqref{eq:unst-ftf-parent2}, respectively, but
we have
$\gamma<0$ for FTW.  As we shall see, however, the properties of the FTW
case cannot be obtained by trivially changing the sign of $\gamma$
in the FTF results.

Consider first the parent of the form (\ref{eq:mu-unstable-ftf}), with $\gamma<0$. For our present purposes it suffices to mention that for $\gamma'>-\epsilon>0$ monotonicity holds, furthermore,  the upper limit of the support $z^\ast$, where $g(z)$ diverges, is some value below $1/|\gamma|$.
Near $z^\ast$ the argument of the log in (\ref{eq:gz-ftf}),
\begin{equation}
 \phi(z)=e^{\gamma g(z)},
\end{equation}
is analytic, whence we can determine the indices of universality following Sec.\ \ref{sS:iid-lin-parent}. We have to determine the effective index
\begin{equation}
 \gamma_N = \gamma + \epsilon_N = - \frac{g''(b_N)}{g'^2(b_N)} =
\gamma - \frac{\gamma \phi(b_N) \phi''(b_N)}{\phi'^2(b_N)}.
\label{eq:unst-ftw1-gammaN}
\end{equation}
Now the $N\to\infty$ limit of the effective $\gamma_N$ gives $\tilde\gamma$,
while the decay of $\epsilon_N$ provides the index $\tilde\gamma'$.
They can be obtained by noting that from \eqref{eq:um-ftf-b} we have $\phi(b_N)=N^\gamma$ and higher derivatives of $\phi(b_N)$ go to constants.  Thus the limit of \eqref{eq:unst-ftw1-gammaN} is
\begin{equation}
 \tilde\gamma=\gamma, \label{eq:unst-ftw1gamma}
\end{equation}
and the exponent of convergence of $\epsilon_N$ is equally
\begin{equation}
 \tilde\gamma'=\gamma.\label{eq:unst-ftw1gammaprime}
\end{equation}
Thus, we found that the unstable manifold returns
to the same fixed point along the stable manifold whose stability
parameter $\tilde\gamma'$ coincides with the fixed point parameter,
irrespective of the initial value of the instability parameter $\gamma'$.

Turning to the case of the additively perturbed distribution, we use  \eqref{eq:unst-ftf-parent2} with $\gamma<0$.
If $1>-\gamma>\gamma'>0$ then, for not too large $\epsilon>0$,
one has the same upper limit of support $z^\ast=-1/\gamma$
as in the unperturbed case, and
 \begin{equation}
\begin{split}
g(z) &= -\ln (-\ln \mu(z) ) \\ &\approx \frac{1+\gamma}{\gamma}
\ln(1+\gamma z) + d_1 \\ &+ d_2(1+\gamma z)^{\gamma'/\gamma+1} +d_3(1+\gamma
z)^{-(1+\gamma)/\gamma},
 \end{split} \label{eq:unst-ftw-g2}
\end{equation}
where $d_j$ are constants.  We calculate again
the effective $\gamma_N$ according to Sec.\ \ref{sS:iid-lin-parent},
and wind up with the following final indices
\begin{subequations}
\begin{align}
\tilde\gamma &= \dfrac{\gamma}{1+\gamma}, \\
 \tilde \gamma' &= \max \left\lbrace  \frac{\gamma+\gamma'}{1+\gamma} , -1 \right\rbrace .
\end{align}
\label{eq:gammatildes-ftw2}
\end{subequations}
Comparison of \eqref{eq:gammatildes-ftw2} with (\ref{eq:unst-ftw1gamma},\ref{eq:unst-ftw1gammaprime}) demonstrates once more that the asymptotic properties strongly depend on the way the unstable perturbation is applied. 
Plots of the evolution of distributions as $N$ increases 
for initial states with $\gamma<0<\gamma'$ are qualitatively similar
to previous figures, therefore we do not display them.

%%%%%%%%%%%%%%%%%%%%%%%%%%%%%%%%%%%%%%%%%%%%%%%%%%%%%%%%%%%%%%%%%%%%%%%%%%%%
\subsection{Unstable manifolds in function space}
\label{S:iid-unst-fig}
%%%%%%%%%%%%%%%%%%%%%%%%%%%%%%%%%%%%%%%%%%%%%%%%%%%%%%%%%%%%%%%%%%%%%%%%%%%%

\begin{figure}[htb]
\psfrag{c}{$\gamma>0$}
\psfrag{b}{$\tilde\gamma(<\gamma)$}
\psfrag{f}{$\gamma'\!>\!0$}
\psfrag{a}{$\tilde\gamma'=\tilde\gamma=\gamma=0$}
\psfrag{g}{$\tilde\gamma'\!\!<\!0$}
\psfrag{d}{$\gamma$}
\psfrag{e}{$\tilde\gamma'=\tilde\gamma=\gamma\!<\!0$}
\psfrag{ww}{FTW}
\psfrag{gg}{FTG}
\psfrag{ff}{FTF}
\includegraphics[width=8.3cm]{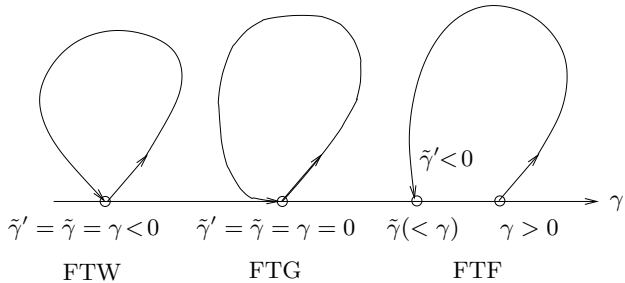}
\caption{Illustration of the invariant manifolds of various fixed
points in the space of probability distributions. Starting
distributions near a fixed point with some
$\gamma$, differing from it by an eigenfunction with a parameter $\gamma'>0$,
with the definition in \eqref{eq:unst-parent1}.  The straight line is the
$\gamma$ axis and the directed curves represent  RG transformation flows.
Each flow ends up in a fixed point with $\tilde\gamma$ along a stable
direction characterized by the finite-size index $\tilde\gamma'\leq 0$.  The
approach to the FTG fixed point is along the marginal direction
$\tilde\gamma'=0$, which
corresponds to a small perturbation on the fixed line, as discussed in Sec.\
\ref{S:iid-lin-mg}.
\label{F:manifolds1}}
\end{figure}

The results above on the unstable manifolds emanating from the eigendirections
of the form \eqref{eq:unst-parent1} are graphically represented in function
space on Fig.\ \ref{F:manifolds1}.  If we start out from a parent distribution
near the fixed point with $\gamma$, in the direction of the unstable
eigenfunction with $\gamma'>0$, then upon the action
of the RG transformation the distribution function moves on an unstable
manifold.  For $\gamma\leq 0$, the FTW case, the unstable manifold loops
back towards the same fixed point, whose neighborhood it started out from.
On the other hand, in the FTF case, $\gamma>0$, the RG trajectory will wind up
in the fixed point with $\tilde\gamma<\gamma$, as given in
\eqref{eq:gammatilde-ftf},  falling between the starting
and the FTG fixed points.  The characteristic parameter of the stability is
$\tilde\gamma'=\gamma$ for the case $\gamma\leq 0$, whereas for the FTF class,
$\gamma>0$, the parameter of the stable eigendirection is given by Eq.\
(\ref{eq:gammaprimetilde-ftf}).

\begin{figure}[htb]
\psfrag{b}{$\gamma>0$}
\psfrag{c}{$\tilde\gamma(>\gamma)$}
\psfrag{f}{$\gamma'>0$}
\psfrag{h}{$\gamma'<0$}
\psfrag{a}{$\tilde\gamma=\gamma=0$}
\psfrag{g}{$\gamma<0$}
\psfrag{d}{$\gamma$}
\psfrag{e}{$\tilde\gamma(<\gamma)$}
\psfrag{ww}{FTW}
\psfrag{gg}{FTG}
\psfrag{ff}{FTF}
\includegraphics[width=8.3cm]{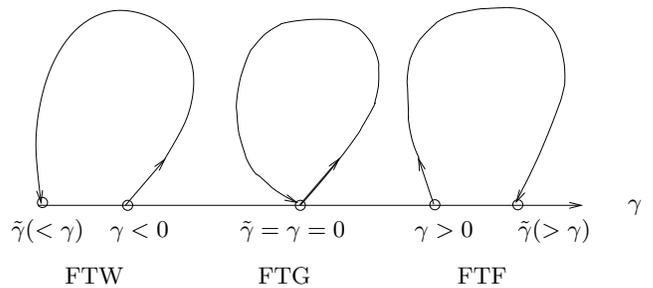}
\caption{Illustration of the invariant manifolds from starting
distributions with $\gamma'>0$ in the definition \eqref{eq:unst-parent2}.
Each flow converges to a fixed point with $\tilde\gamma$ along a stable
direction characterized by a finite-size index $\tilde\gamma'<0$.
\label{F:manifolds2}}
\end{figure}

The invariant manifolds starting from unstable distributions of the form
\eqref{eq:unst-parent2} are illustrated on Fig.\ \ref{F:manifolds2}.  For the
cases $\gamma\ne 0$ the RG trajectory will wind up in the fixed point with a
different $\tilde\gamma$, farther from the origin. This can be understood from
the fact that the additive perturbation in \eqref{eq:unst-parent2} slows down
the asymptote of the distribution, while such effect is not present if the perturbation is in the argument like in \eqref{eq:unst-parent1}.   The characteristic parameter of the
asymptotic stability is $\tilde\gamma'<0$ in all cases.

Note that on the fixed line there is a symmetry between a $\gamma>0$
and its negation, i.e., the FTF and FTW cases. In particular, in the
FTF case with a $\gamma>0$, the variable $1+\gamma
x$ has the same distribution as in the FTW case, with the parameter $-\gamma$, the variable $1/(1-\gamma x)$ has. Nevertheless, this symmetry does not manifest itself for the invariant manifolds. On the one hand, in  \eqref{eq:unst-parent1} the symmetry is broken, since $\psi$ is added to the variable itself in
the argument.  This is also responsible for the effect that in the FTF
case the decay of the limit distribution is stronger than that of the
parent.  On the other hand, for the additively perturbed distributions
of the form \eqref{eq:unst-parent2} there is a qualitative symmetry,
in that the RG trajectory loops back to the fixed line farther from
the origin for both signs of the starting $\gamma$.   The particular values of the final $\tilde\gamma,\tilde\gamma'$ indices, as given by \eqref{eq:gammatildes-ftf2} and \eqref{eq:gammatildes-ftw2}, however, do not exhibit the sign flip symmetry of the fixed line.

%%%%%%%%%%%%%%%%%%%%%%%%%%%%%%%%%%%%%%%%%%%%%%%%%%%%%%%%%%%%
\section{Conclusion and outlook}
\label{S:end}
%%%%%%%%%%%%%%%%%%%%%%%%%%%%%%%%%%%%%%%%%%%%%%%%%%%%%%%%%%%%%%%%%%

We have shown how finite-size corrections to EVS limit distributions
can be derived by borrowing ideas and terminology from the
RG theory. The corrections in shape and amplitude emerged from the
action of the RG transformation on
the function space associated with the close neighborhood of
the fixed line of limit distributions. This RG approach
provided the classification of the
first-order corrections and,
furthermore, it also helped to establish the connection
to parent distributions, thus making the results practically useful.
A remarkable feature of the RG theory is that
the study of the linear neighborhood of the fixed line reveals
orbits -- invariant manifolds -- of the RG transformation which first
diverge and later return to the same or to a different
point on the fixed line.

There are several problems in EVS where the RG approach developed
here may be of interest. Within the theory of EVS of i.i.d.\ variables,
the RG theory should be instrumental in developing a compact
treatment of the finite-size corrections for order statistics in
general, among them for the 2nd, 3rd, etc.\ maxima, as well as
for their various joint
statistics. Furthermore, higher order corrections are also of
interest from the empirical viewpoint,
especially in the case of the logarithmically slow convergence.

The next level of difficulty is met when trying to investigate
the problem of weakly correlated variables (a nontrivial 
example is the stationary sequence of Gaussian variables 
with correlation decaying as a power law) where the EVS limit 
distribution of the maximum is
known to be FTG \cite{Berman:1964}). There the
question arises naturally, whether the finite-size corrections
elucidated in this paper remain valid
for families of sufficiently weakly correlated variables.
There is ground to believe that the finite-size corrections
may be the same \cite{GyorgyiETAL:2007} but further
studies are needed to clarify the matter.

Finally, the extension of the RG approach to calculate the EVS
of strongly correlated
systems appears to be a real challenge. We are aware of only
one work \cite{SchehrLeDoussal:2009}
where the so called real-space renormalization group
was developed for obtaining EVS of random walk type processes.
The results are restricted presently to the limit distributions
for these processes. It would be certainly interesting to extend
the method to find the finite-size corrections and compare with
the phenomenological approaches \cite{GyorgyiETAL:2008} since 
some of them suggests
that the shape corrections can be calculated from the
limit distributions.

\vspace{5pt}

\begin{acknowledgments}
This research has been supported by the Swiss NSF and by the
Hungarian Academy of Sciences through OTKA Grants
Nos.\ K 68109, K 75324 and NK 72037. 
MD acknowledges the hospitality
of the Physics Institute of the E\"otv\"os University.
\end{acknowledgments}

%\bibliography{references_evs_v4}

\end{document}